\pdfoutput=1

\documentclass[9pt, conference]{IEEEtran}
\IEEEoverridecommandlockouts
\usepackage{cite}
\usepackage{amsmath,amssymb,amsfonts}
\usepackage{algorithmic}
\usepackage{textcomp}

\usepackage{setspace}
\def\smallerspacecaption{\vspace{-2mm}}

\usepackage{floatrow}
\usepackage[caption=false,font=footnotesize]{subfig}

\usepackage[table, dvipsnames]{xcolor}  
\usepackage{fancyhdr}
\usepackage{graphicx}
\usepackage{url}
\usepackage{verbatim}
\usepackage{multirow}
\usepackage{hhline}
\usepackage[us]{datetime}
\usepackage{paralist}
\usepackage{stfloats}
\usepackage[ruled, vlined, norelsize]{algorithm2e} 
\SetAlFnt{\sf} 

\graphicspath{{figures/}}

\usepackage{soul}
\newcommand{\drop}[1]{\textcolor{blue}{\st{#1}}}
\renewcommand{\drop}[1]{}

\newdimen\arrayruleHwidth
\makeatletter
\def\Hline{\noalign{\ifnum0=`}\fi\hrule \@height \arrayruleHwidth
\futurelet \@tempa\@xhline}
\makeatother

\makeatletter
\def\blfootnote{\xdef\@thefnmark{}\@footnotetext}
\makeatother

\shortdate
\settimeformat{ampmtime}

\makeatletter
\newcommand*\tabsize{%
	   \@setfontsize\tabsize{6}{7.2}%
}
\makeatother

\begin{document}

\title{Rethinking Split Manufacturing: An Information-Theoretic Approach with Secure Layout Techniques\\
}
%
%
%
%
%


%
\newcommand*\samethanks[1][\value{footnote}]{\footnotemark[#1]}

\author{
	{\Large Abhrajit Sengupta$^{\dagger}$\textsuperscript{*}\thanks{\textsuperscript{*}A.\ Sengupta and S.\ Patnaik contributed equally.}, Satwik Patnaik$^{\dagger}$\textsuperscript{*}, Johann Knechtel$^{\ddagger}$, Mohammed Ashraf\,$^{\ddagger}$,
	   Siddharth Garg$^{\dagger}$,}\\{\Large and Ozgur Sinanoglu$^{\ddagger}$}\\[2pt]
  {\large $^{\dagger}$\,Tandon School of Engineering, New York University, New York, USA}\\
  {\large $^{\ddagger}$\,New York University Abu Dhabi, Abu Dhabi, United Arab Emirates}\\
  \normalsize{\{as9397, sp4012, johann, ma199, sg175, ozgursin\}@nyu.edu}
}


\maketitle

\renewcommand{\headrulewidth}{0.0pt}
\thispagestyle{fancy}
\pagestyle{fancy}
\cfoot{
	\vspace{-1cm}
\copyright~2017 IEEE.
		Personal use of this material is permitted. Permission from IEEE must be obtained for all other uses, in any current or future media, including
		reprinting/republishing this material for advertising or promotional purposes, creating new collective works, for resale or redistribution to servers or lists, or
		reuse of any copyrighted component of this work in other works.\\
	The definitive Version of Record is published in
	Proc. International Conference On Computer Aided Design (ICCAD) 2017\\
http://dx.doi.org/10.1109/ICCAD.2017.8203796
}

\begin{abstract}

Split manufacturing is a promising technique to defend against fab-based malicious activities such as IP piracy, overbuilding, and insertion of hardware Trojans. 
However, a 
network flow-based
proximity attack,
proposed by Wang \emph{et al.\ }(DAC'16)~\cite{jv-attack16},
has demonstrated that most prior
art on split manufacturing is highly vulnerable.
Here in this work, we present two
practical layout techniques towards secure split manufacturing:
(i) gate-level graph coloring and (ii) clustering of same-type gates.
Our approach shows promising results against the advanced proximity attack,
	lowering its success rate by 5.27\emph{x},
3.19\emph{x}, and 1.73\emph{x} on average compared to the unprotected layouts when 
splitting at metal layers M1, M2, and M3, respectively. 
Also, it largely outperforms previous defense efforts; we observe on average 8\emph{x} higher resilience when compared to representative prior art. 
At the same time, extensive simulations on ISCAS'85 and MCNC
benchmarks reveal that our techniques incur an acceptable layout overhead. 
Apart from this empirical study, we provide---for the first time---a theoretical framework for quantifying the
layout-level resilience against any proximity-induced information leakage.  Towards
this end, we leverage the notion of mutual information and provide
extensive results to validate our model.

\end{abstract}

\section{Introduction} \label{sec:intro}
Nowadays, more and more
companies rely on
external foundries for cost-effective access to advanced
fabrication technologies.
However, as this trend towards globalization of integrated circuit (IC) manufacturing consolidates, 
companies are forced to share their valuable intellectual property (IP) with potentially untrusted parties.
This dependency coupled with currently inadequate
protection measures has led to many security vulnerabilities
such as IP piracy, overbuilding, and insertion of hardware Trojans~\cite{trojan1,
	re2, primer14}. 
These threats are becoming an increasing concern for both commercial and military organizations. 
In fact, it is estimated that
several billions of dollars are lost each year owing to IP piracy~\cite{semi}.

\subsection{Split Manufacturing and Proximity Attack}
Split manufacturing was proposed by the IARPA
agency~\cite{iarpa} to thwart 
the aforementioned threats.
Leveraging the asymmetry of the metal layers, the design is split into two parts: the front-end-of-line (FEOL), consisting of the active device layer and lower metal layers (e.g.,
		$\leq$ M3),
and the back-end-of-line (BEOL), that is the remaining higher metal layers (e.g., $\geq$ M4).\footnote{In accordance with~\cite{jv-attack16}, our notion of splitting, e.g., at M2,
	implies that metal layers M1 and M2 as well as V23 (i.e., the vias between M2 and M3) are readily available to fab-based attackers.}
   The FEOL is manufactured in a high-end, third-party foundry which is
\emph{untrusted}, whereas the BEOL
is fabricated at a \emph{trusted} facility on top of the incomplete wafer(s) provided by the FEOL foundry.
This two-step approach 
helps to hide the overall functionality of the design
from an
attacker residing at the FEOL
foundry, thereby hindering her/him
from pirating the design or maliciously modifying it via hardware Trojans.
Recently, different works have successfully demonstrated the feasibility of split manufacturing~\cite{sp1, sp2, sp3,otero15}.

Unfortunately, naive split manufacturing falls short of ensuring security.
Commercially available physical-design tools apply certain heuristics to minimize power, performance, and area,
which may leak certain information. 
An attacker in the foundry
can leverage
this information to retrieve the missing BEOL connections, possibly undermining the defense intended by split manufacturing.
In fact, Rajendran \emph{et al.}~\cite{jv-attack13}
exploit
the physical proximity between the cells to be connected;
they
demonstrated a \emph{proximity attack} that connects
nearby
cells to retrieve the missing BEOL connections. 
Recently, an advanced network-flow attack was presented by Wang \emph{et al.}~\cite{jv-attack16}---this attack has been shown to
render most prior protection schemes for split manufacturing insecure.

\begin{figure*}[tb]
\centering
\includegraphics[width=0.85\textwidth]{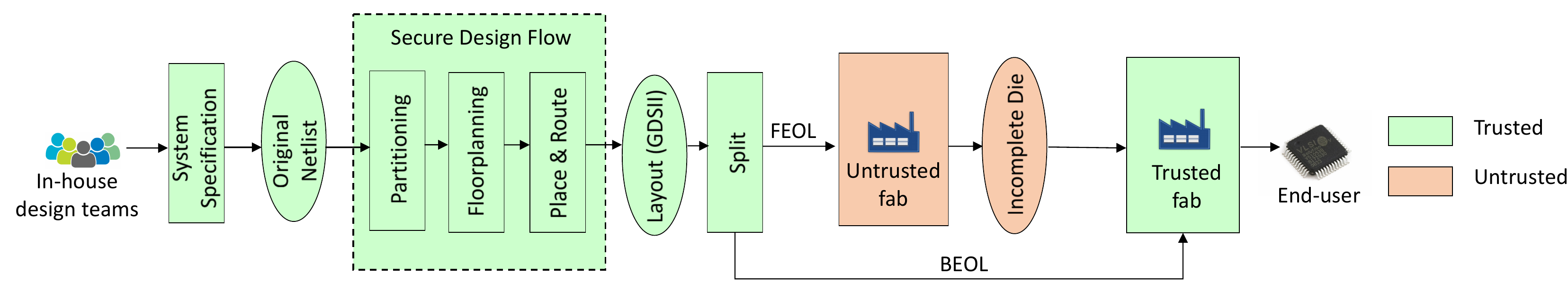} 
\smallerspacecaption
\caption{Threat model for split manufacturing, along with our secure design flow (dashed).
	Note that the untrusted FEOL fab may want to pirate some IP and/or insert hardware Trojans. In this work, we primarily focus on the former aspect of the threat model.
}
\label{fig:threat_model}
\end{figure*}

The threat model for split manufacturing is depicted in Fig.~\ref{fig:threat_model}.
The attacker has access to the technology libraries 
but is oblivious of the functionality of the IC. Naturally, she/he also cannot obtain a working IC; the IC is yet to be manufactured.

\subsection{Prior Art and Our Contributions}
To thwart proximity attacks, a pin-swapping based countermeasure was proposed in~\cite{jv-attack13}.
In practice, however, many swapped connections can still be correctly inferred.
Jagasivamini \emph{et al.}~\cite{jaggu14}
showed that splitting at
a lower layer renders the design more secure against proximity attacks; they propose to split the design at M1.
Though splitting at M1 may render the design secure against such attacks, it also necessitates state-of-the-art manufacturing facilities
at the trusted BEOL foundry. As a result, the cost of production significantly increases, defeating one of the promises of split manufacturing, i.e., affordable (but secure)
		IC production~\cite{sp1, sp2, sp3}.
Besides, Wang \emph{et al.}~\cite{jv-attack16} proposed an algorithm for heuristic placement perturbation towards layout protection.
Wang \emph{et al.}~\cite{wang17} further proposed
a routing-based scheme
targeting for 50\% Hamming distance
between the original and the reconstructed netlist
(to induce the maximal ambiguity for an attacker).
Maga\~{n}a \emph{et al.}~\cite{magana16} insert routing blockages to lift wires and, thus, to mitigate routing-centric attacks.
To counter Trojan insertion, a
formal method for ``k-security by wire lifting'' was proposed 
by Imeson \emph{et al.}~\cite{imeson13};
	it comes along with a high overhead, e.g., $\approx200\%$ for delay.

Apart from this, there has been little to no effort towards
a theoretical model which can quantify the resilience of a design against proximity attacks in general. In this work, we
propose such a model based on the concepts of entropy and mutual information. Although the notion of entropy was previously advocated by Jagasivamani \emph{et al.}~\cite{jaggu14},
	their study lacks
specific formulations and, thus, fails to measure
resilience 
in both theory and practice.

Building up on our theoretical framework, we propose several \emph{placement-centric} techniques aiming to make split manufacturing secure
against any proximity attack while ensuring practicality. As for
our baseline approach, i.e., full randomization of the placement, it provides the highest
level of security, but also incurs the highest layout overhead. Thus, to reduce overhead, we propose two novel 
techniques based on graph coloring and clustering gates of the same type. We show empirically that these techniques can attain
notably better trade-offs for layout cost and security.

The contributions of our work can be summarized as follows:
\begin{compactitem}
\item An information-theoretic framework to gauge the resilience of a given layout against any
	proximity attack (Section~\ref{sec:theory}).
\item Two placement-centric techniques which help to render split manufacturing-based layouts secure against any proximity attack at acceptable overhead (Section~\ref{sec:techniques}).
\item A thorough investigation based on the well-known ISCAS'85 and MCNC benchmarks, demonstrating the effectiveness of our techniques and contrasting with naive
randomization and prior art
(Section~\ref{sec:results}). Here we also investigate the cost-security trade-offs for split manufacturing induced by different split layers; we look into splitting at M1 up to M6.
\end{compactitem}

\section{Information-Theoretic Metric}\label{sec:theory}
Recall that split manufacturing is meant to offer protection against fab-based attacks such as
IP piracy, overbuilding, and/or insertion of hardware Trojans. 
While the intended protection is based on the fact that
the FEOL and BEOL
of the chip are manufactured by different parties,
      physical design tools still operate on the entire design holistically, driven by the strong need for design and cost optimization.
As a result, any partial, FEOL-level layout might leak certain information which can be leveraged by an attacker to infer the hidden BEOL connections.
Indeed,
the notion of physical proximity between connected cells, among other hints, has been leveraged in multiple attacks~\cite{jv-attack13,jv-attack16,magana16}.

An attack-based, empirical security evaluation
has two major drawbacks: (1) it can be
time-consuming and, thus, ineffective for large layouts; (2)
it is
naturally specific to the employed attack and, thus, fails to quantify
the layout's protection (or the lack of) 
against other attacks.
Surprisingly, however, there has been very little
to no effort towards measuring the resilience of a layout against an advanced or even an optimal attack.
In this regard, we introduce \emph{an information-theoretic metric
		to quantify
the amount of information that can be extracted by an attacker from the physical layout}.

\subsection{Measures of Information Leakage}
To measure the uncertainty of an attacker about the missing connectivity of a given layout, we leverage
the concept of \emph{entropy}, which was famously introduced by Shannon.
Note that the concept of entropy has been extensively employed to assess the vulnerability 
of cryptosystems in the context of side-channel attacks, such as power analysis or timing attacks~\cite{kopf07, stan09}.

The entropy of a
 variable $X:\mathbb{X}$ is defined as  
\begin{equation}
H[X] = - \sum_{x\in X}^{} \mbox{Pr}[X=x] \cdot \mbox{log Pr}[X=x]
\end{equation}
Given another
variable $Y:\mathbb{Y}$, the conditional entropy of $X$ denoted as $H[X|Y]$ can be
expressed as
\begin{equation}
H[X|Y] = - \sum_{y\in Y}^{} \mbox{Pr}[Y=y] \cdot H[X|Y=y]
\end{equation}
The attacker's initial uncertainty about $X$ is $H[X]$, and given a leakage model denoted
by $Y$, the amount of information leakage---formally termed as \emph{mutual information  (MI)}~\cite{stan09}---is expressed as
\begin{equation}\label{eqn:mi}
I(X;Y) = H[X] - H[X|Y] 
\end{equation}
\vspace{-1cm}

\subsection{Resilience Against Proximity Attacks}
Any proximity attack leverages the fact that the distance between cells reveals information
about their connectivity. Thus, by analogy, the distance between cells constitutes the leakage
model which an attacker tries to exploit.
	Hence,
	we define two variables, $X$ and $D$, capturing the connectivity and distance between cells as 
\begin{align}
    X &= 
\begin{cases}
    1& \text{if two cells $u$ and $v$ are connected;}\\
    0              & \text{otherwise}
\end{cases}\\
D &= \mbox{\texttt{distance}}(u,v)
\end{align}
Without loss of generality, we apply the notion of Manhattan distance (sum of horizontal and vertical distance) between cells.

To quantify the amount of information revealed by their distance about the connectivity between cells and, thus,
to quantify the resilience of a layout against proximity attacks,
we determine the mutual information \emph{MI}
\begin{align}
MI = I(X;D) = H[X] - H[X|D]
\end{align}
Note that the conditional entropy $H[X|D]$ itself can serve a similar purpose, but it fails to capture the notion of information leakage.

To compute $H[X]$ and $H[X|D]$, we determine the distribution of $X$ and $D$ for a given layout
in a pairwise manner for all gates, allowing a straightforward and efficient computation of $I(X;D)$.

\begin{figure}[tb]
\centering
\includegraphics[width=0.645\textwidth]{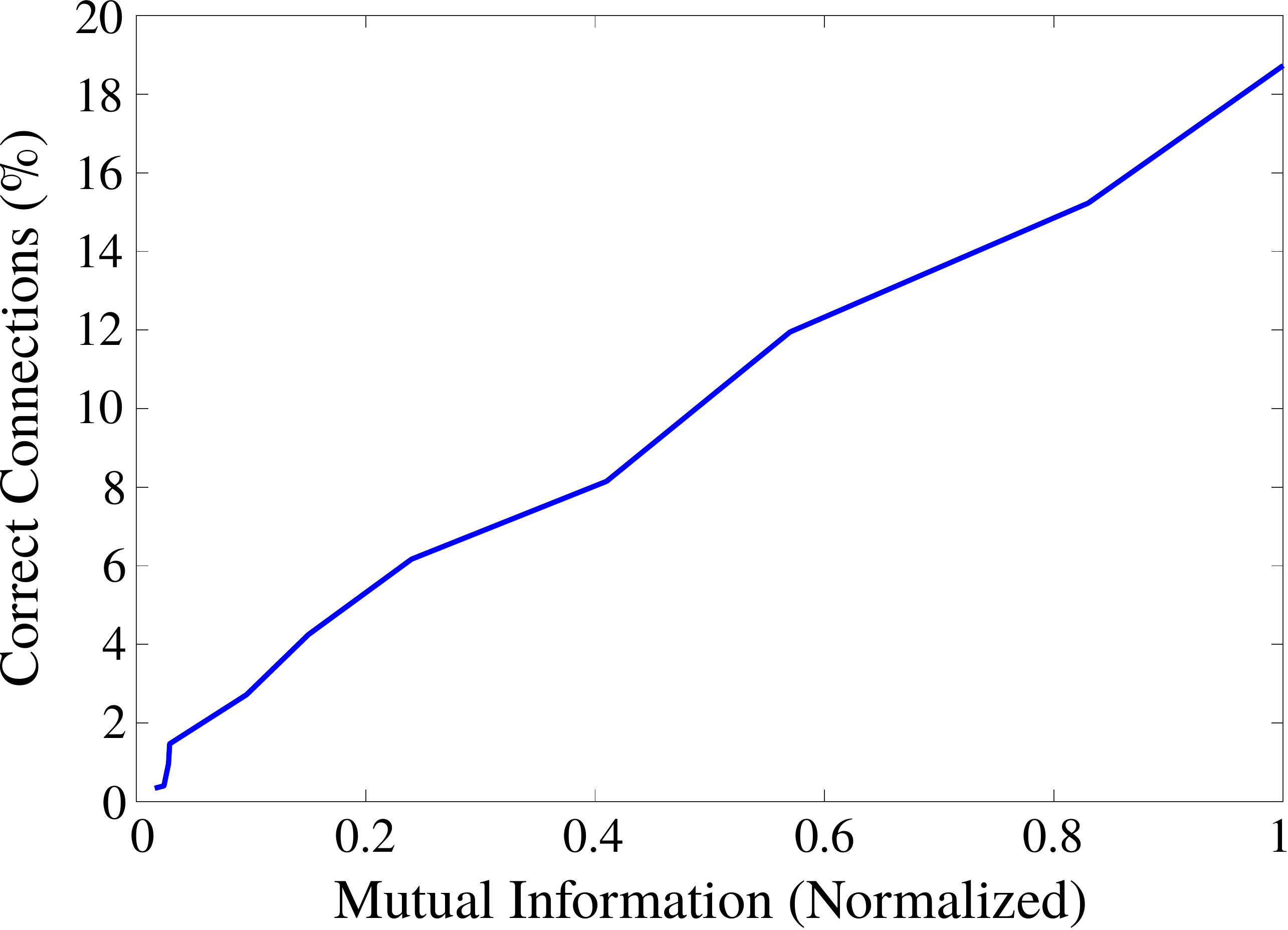} 
\smallerspacecaption
\caption{Correct connections over normalized mutual information for stepwise randomized layouts of \emph{c7552}, split at M1.}
\label{fig:mi_vs_cc}
\end{figure}

The \emph{MI} quantifies the inherent 
protection of a layout against proximity attacks; the lower the \emph{MI}, the lower the correlation
between connectivity $X$ and distance $D$ and, thus, the better the protection. 
This correlation is apparent from Fig.~\ref{fig:mi_vs_cc} where the graph for
correctly recovered connections (by running the proximity attack of~\cite{jv-attack16}) is plotted
over the normalized \emph{MI} for the \emph{c7552} benchmark split at M1.
Here
	we shuffle the placement of randomly selected cells (from 0 to 100\% of all cells, in steps of 10\%).
This way,
     we obtain 11 different layouts with varying and unbiased distributions for the \emph{MI}.
The plot reveals a linear relation between the \emph{MI} and the correct connections (i.e., the attacker's success rate),
validating our hypothesis that a lower \emph{MI} implies higher security.

The goal of a security-aware designer is thus to generate layouts
in such a way that the \emph{MI} is minimized.
Also, another interesting measure could be $I(D;X)$, i.e., the amount of information revealed by the connectivity about the distance of gates, but it turns out that
\begin{align*}
I(D;X) &= H[D] - H[D|X]\\
&= H[D] - ([H[X|D] - H[X] + H[D])\\
&= H[X] - H[X|D] = I(X;D)
\end{align*}

So far, we have considered the distance between cells; thus, the \emph{MI}
only quantifies the layout's security when split at M1. 
Such findings may not translate well for splitting at higher metal layers (see also Section~\ref{sec:results}). 
Nonetheless, the proposed information-theoretic metric is generic in the sense that 
it is still applicable to higher split layers as well---one has to simply consider
the distances of \emph{open pins/vias}
rather than the distances between cells.\footnote{Open pins/vias describe the open ends of ``dangling wires'' remaining in the FEOL after split manufacturing; see also~\cite{wang17}.}
Since these pins/vias represents parts of the overall routing infrastructure, which is typically optimized towards short interconnects,
they will leak additional information beyond the placement of cells.

Notably, applying our metric
at different layers will
guide the designer
which layer he/she should split at, using a precise and quantified
	trade-off between security (higher when split at lower layers) and cost
	(lower when split at higher layers).
     As we focus on placement-centric techniques, here we 
compute the \emph{MI} considering gate distances---our metric readily and accurately evaluates the layout protection (or lack of) when splitting at M1.

\section{Our Secure Layout Techniques}\label{sec:techniques}
   Next,
   we present different placement-centric techniques for making a layout secure in the context of split manufacturing and proximity attacks.
Our analysis above elucidates the need to minimize the layout's mutual information (\emph{MI}) of connectivity and distances, to mitigate any proximity attack.

\begin{figure}[tb]
\centering
\includegraphics[width=0.665\textwidth]{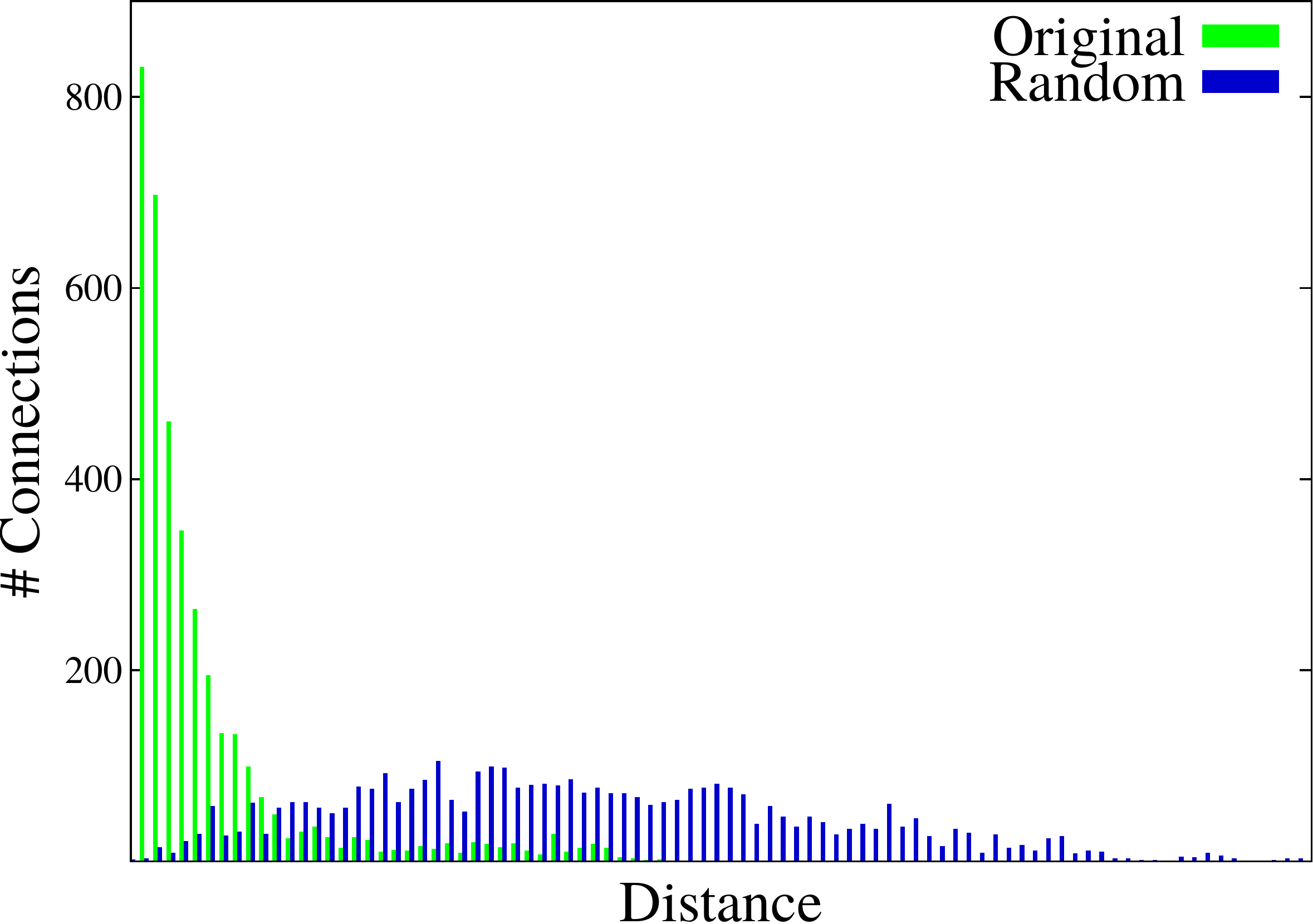} 
\smallerspacecaption
\caption{Distribution of connectivity over distance of \emph{c7552} for original (green) and randomized layout (blue).}
\label{fig:hist_comb}
\end{figure}

\begin{figure}[b]
\centering
\includegraphics[width=0.79\textwidth]{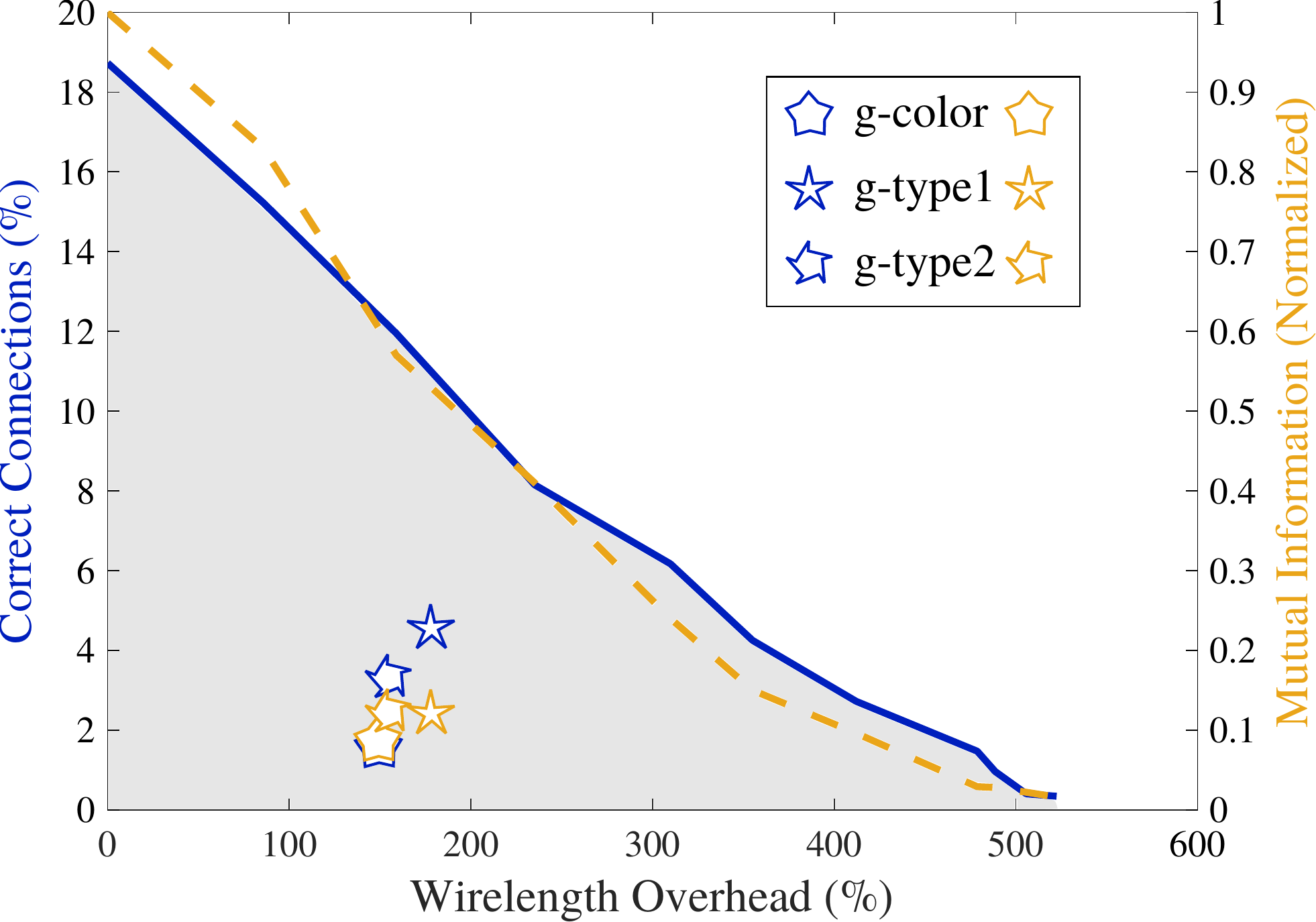} 
\smallerspacecaption
\caption{Correct connections and mutual information versus wirelength overheads, when splitting \emph{c7552} at M1. For layout randomization (represented by the blue line and the dark-yellow, dashed line), gates 
	are randomly selected
	    in steps of 10\% and have their locations shuffled.}
\label{fig:cc_mi_wl}
\end{figure}

One obvious and straightforward (thus naive) idea is to completely randomize the placement of 
cells in the layout to achieve the desired effect.
The intuition here
is that randomizing
a layout 
would stretch the connected cells apart in an unpredictable manner, thus successfully eliminating
any proximity-induced information leakage. 
This is illustrated in Fig.~\ref{fig:hist_comb} where the distribution of connectivity
is plotted against distance for the original and randomized layout of \emph{c7552}, respectively.
It is easy to see
that the connectivity in the randomized layout is nearly uniformly 
distributed over the distance, unlike the original one which is heavily correlated with distance.
The random layout exhibits a very low \emph{MI}, and
is expected to be secure even against advanced proximity attacks.
However, it also incurs excessive overhead regarding power, performance, area, and wirelength, sometimes up to 600\%.
In Fig.~\ref{fig:cc_mi_wl}, for example, we plot the correct connections and \emph{MI} against wirelength overhead for \emph{c7552} when split at M1.
The grey-shaded region in the plot marks the desirable solution space
having better trade-offs for security
and layout cost when compared to randomization.
This raises the following question: \emph{can we develop layout techniques that may approximate or even improve the security/resilience level of layout randomization yet
	at a reasonable cost?}

Here we take on this challenge and present two novel layout techniques, called \emph{g-color} and \emph{g-type}.
As illustrated in Fig.~\ref{fig:cc_mi_wl}, our techniques
can achieve a similar level of security when compared to randomization, with much lower
wirelength overheads at the same time (see also Section~\ref{sec:ppa} for more details on layout cost).
In the next two subsections, we present our techniques.

\subsection{g-color}
We leverage
\emph{graph coloring} 
to hide the connectivity information; coloring
a netlist mandates that there be no connectivity between gates of the same color.
The ``colored
netlist'' is then partitioned by clustering all cells of same colors together and the placement of cells is confined within their respective clusters.
These constraints naturally mitigate the
information leakage to a great extent, thereby
making the layout more secure,
albeit
in a cost-effective manner.

\begin{figure}[tb]
\centering
\includegraphics[width=0.6\textwidth]{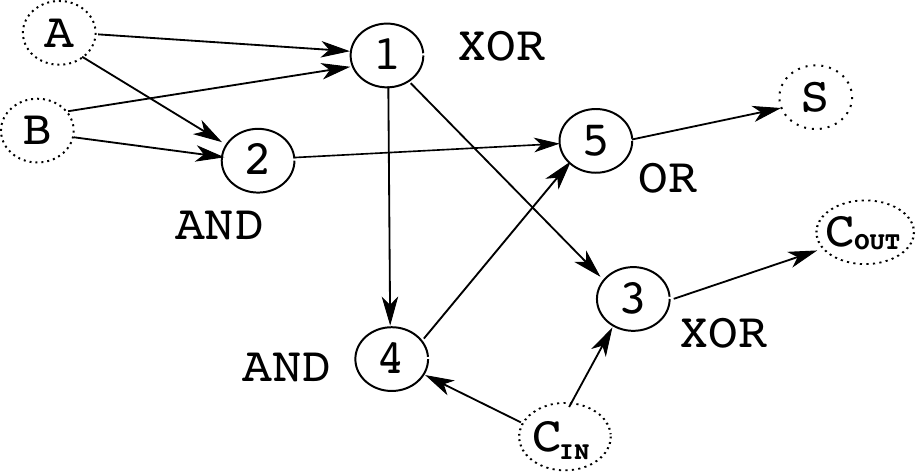}
\smallerspacecaption
\caption{Graph representation of a full-adder circuit.}
\label{fig:fa}
\end{figure}

\begin{figure}[tb]
\centering
\subfloat[Naive, unbalanced coloring.]{
\includegraphics[width=.976\textwidth]{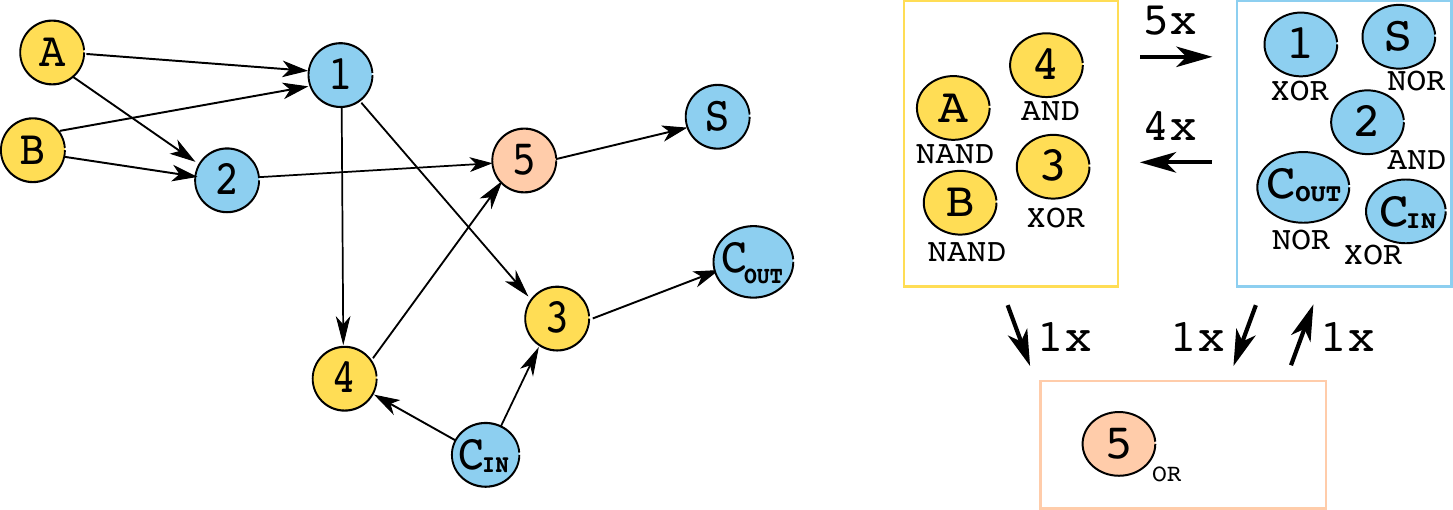}
\label{fig:g-color1}}\\
\subfloat[Balanced coloring, also with different colors for neighbours.]{
\includegraphics[width=.976\textwidth]{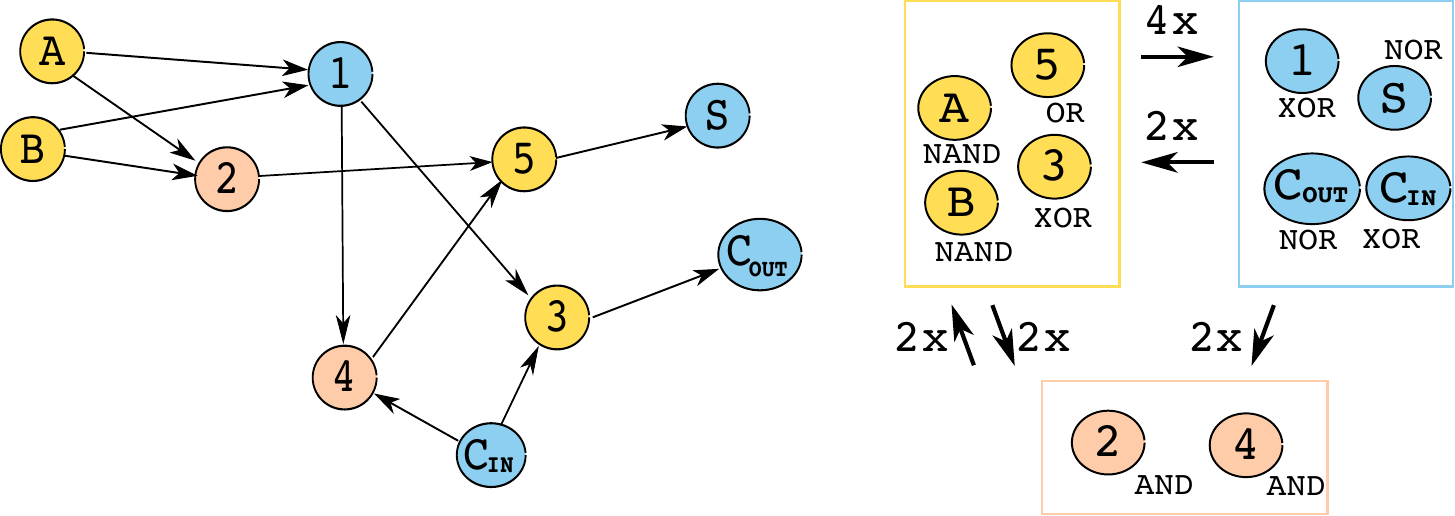}
\label{fig:g-color2}}
\smallerspacecaption
\caption{Applying g-color to a full adder, with the resulting design partitions and their system-level connectivity (right).}\label{fig:g-color}
\end{figure}

The coloring technique is described in Algorithm~\ref{algo:g-color}.
(The reader is also referred to Section~\ref{sec:method}
for further details on layout generation.)
We extend the greedy coloring strategy discussed in~\cite{west}.
The process is illustrated in Fig.~\ref{fig:g-color} where we show the coloring of a full-adder circuit (see also Fig.~\ref{fig:fa} for the latter).
For the sake of simplicity, the inputs and outputs are also considered as vertices/gates, as they are likely connecting to other cells in the overall design. 
The first vertex
	is selected at random and the rest of the vertices are colored iteratively.\footnote{Note that this random selection of the first vertex allows us to obtain different
		versions of protected layouts for the same design.}
After coloring all the vertices/gates, they are clustered together according to their colors as indicated by the encapsulating boxes in Fig.~\ref{fig:g-color}.

\begin{algorithm}[tb]
\tiny
    \SetKwInOut{Input}{Input}
    \SetKwInOut{Output}{Output}

    \Input{Flattened netlist $N$ }
    \Output{Partitioned netlist $N'$}
    $G \leftarrow convertToDAG(N)$ /*convert $N$ to a directed acyclic graph $G$*/\\
    $L \leftarrow getListOfVertices(G)$ /*parse the list of vertices*/\\
    $C \leftarrow \phi$ /*initialize the set of colors*/\\
    \While{$isNotEmpty(L)$}
    {
    $u  \xleftarrow{\$} getNextVertex(L)$ /*pick start vertex randomly*/\\
    \If{$notColorable(u)$}{ 
	$C \leftarrow addNewColor()$ /*requires new color for $u$*/
    }
    $u.color  \leftarrow getMinColor(C)$ /*find the color with fewest cells and color $u$*/ \\
    $L'  \leftarrow getAdjacencyList(u)$ /*list neighbors of $u$*/ \\
    \While{$isNotEmpty(L')$}
    {
    	$v  \leftarrow getNextVertex(L')$ /*color all neighbors of $u$*/\\
    	\If{$notColorable(v)$}{
		$C \leftarrow addNewColor()$
	}
    	$v.color  \leftarrow getMinColor(C)$ \\
	$delete(v, L')$\\
	$delete(v, L)$\\
    }
    $delete(u, L)$\\
    }
    $N' \leftarrow partitionByColor(N,C)$ /*partition the netlist according to color*/ \\ 
    \Return $N'$
    \caption{Algorithm for g-color} 
    \label{algo:g-color}
\end{algorithm}

Initially, we observe that
	selecting the first available color for the next vertex may
produce largely unbalanced clusters (Fig.~\ref{fig:g-color1}).
In turn, this can cause the design tool to place the different types of partitions (small/large, little/largely interconnected with other partitions) in a manner which may leak
information about the underlying connectivity of the gates.
We thus adapt the algorithm to select the color corresponding to that with the so-far lowest number of associated cells,
   yielding more balanced partitions in practice (Fig.~\ref{fig:g-color2}).

So far,
   we select the same color for all the neighbors of a vertex $v$; all the neighbors/cells
   are consequently assigned to the same partition.
   However, as these neighbors/cells are all driven by the same cell (the vertex $v$), any layout tool seeks to place them in close proximity within their partition.
Thus, we further adapt the algorithm to explicitly color all the neighbors differently, thereby ``decoupling'' cells from their driver.\footnote{Conceptually, we now realize
	coloring of a {\em
	hyper-graph}~\cite{KLMH11}.}
The assignment of different colors to neighbors is streamlined with the balance-aware color selection;
see
Fig.~\ref{fig:g-color2} for an example.

\subsection{g-type}
We observe empirically that the connectivity amongst the same type of gates is rather low;
even in case where particular structures such as ``AND trees'' are present, they seldom dominate the overall design.
Thus,
our second approach (independent of g-color) is to cluster and partition all gates of the same type.

\begin{figure}[tb]
\centering
\includegraphics[width=.98\textwidth]{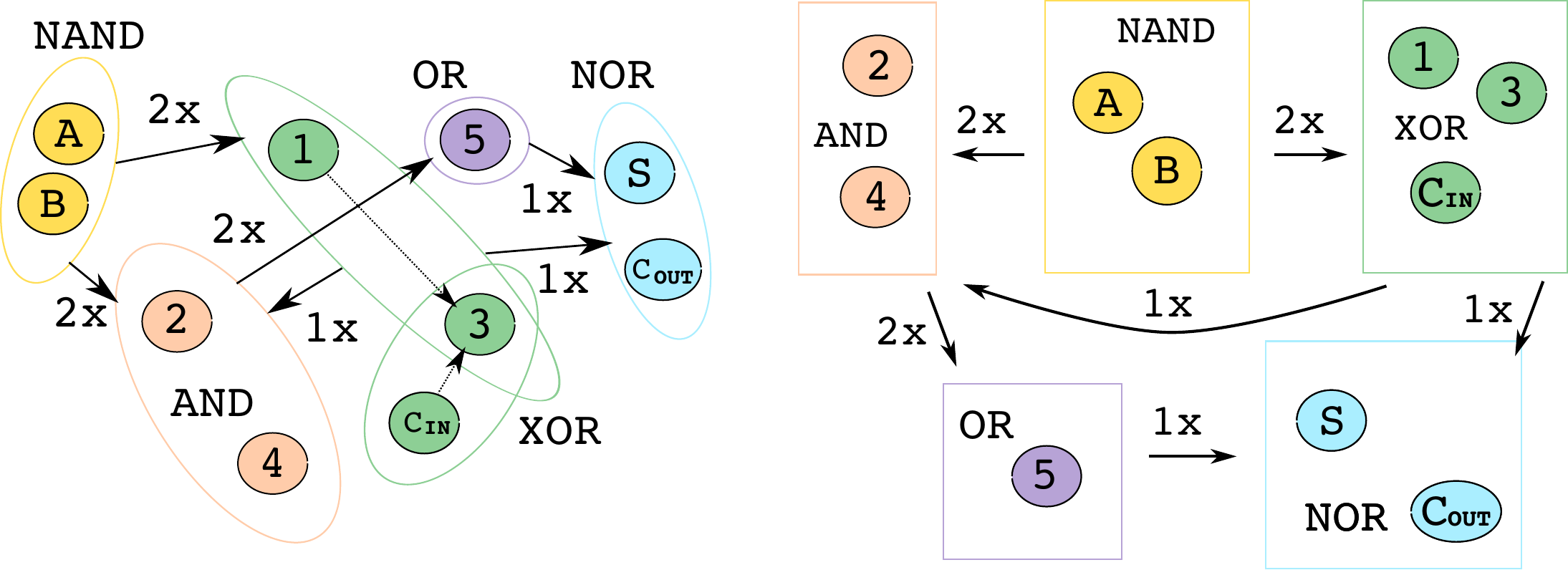}
\smallerspacecaption
\caption{Applying g-type to a full adder, with the partitions.}
\label{fig:g-type}
\end{figure}

Our technique called g-type is outlined in Algorithm~\ref{algo:g-type} and illustrated in Fig.~\ref{fig:g-type}.
It comes in two flavors---we either consider ($i$) only the functionality of the gates  (\emph{g-type1}), or ($ii$)
   both the functionality of the gates as well as the number of their inputs (\emph{g-type2}), e.g. here we do differentiate between a 2-input NAND gate and a 3-input
NAND gate.
The latter is motivated by our
experimental results which indicate
that utilizing more partitions renders a design more resilient against proximity attacks in practice. 
Note that we do not account for driving strengths
during partitioning. Doing so would be superfluous since design tools scale up gates as needed (and/or insert buffers) during later stages.

\begin{algorithm}[tb]
\tiny
    \SetKwInOut{Input}{Input}
    \SetKwInOut{Output}{Output}

    \Input{Flattened netlist $N$ }
    \Output{Partitioned netlist $N'$}
    $L \leftarrow parseNetlist(N)$ /*list the set of vertices*/\\
    $H \leftarrow \phi$ /*initialize gate-types*/\\
    \While{$isNotEmpty(L)$}
    {
    $u  \leftarrow getNextVertex(L)$ \\
    \If{$u.type ==\normalfont{\texttt{BUF or INV}}$}{
	$H \xleftarrow{\$}u$ /*place $u$ uniformly randomly*/
    }
    $H \leftarrow hash(u, u.type)$ /*partition $u$ according to its type*/\\
    $delete(u, L)$\\
    }
    $N' \leftarrow partitionByType(N,H)$ /*partition the netlist by gate-types*/\\ 
    \Return $N'$
    \caption{Algorithm for g-type} 
    \label{algo:g-type}
\end{algorithm}
  
\section{Methodology}\label{sec:method}
\begin{figure}[tb]
 \centering
 \subfloat[g-color]{
 \includegraphics[height=3.85cm]{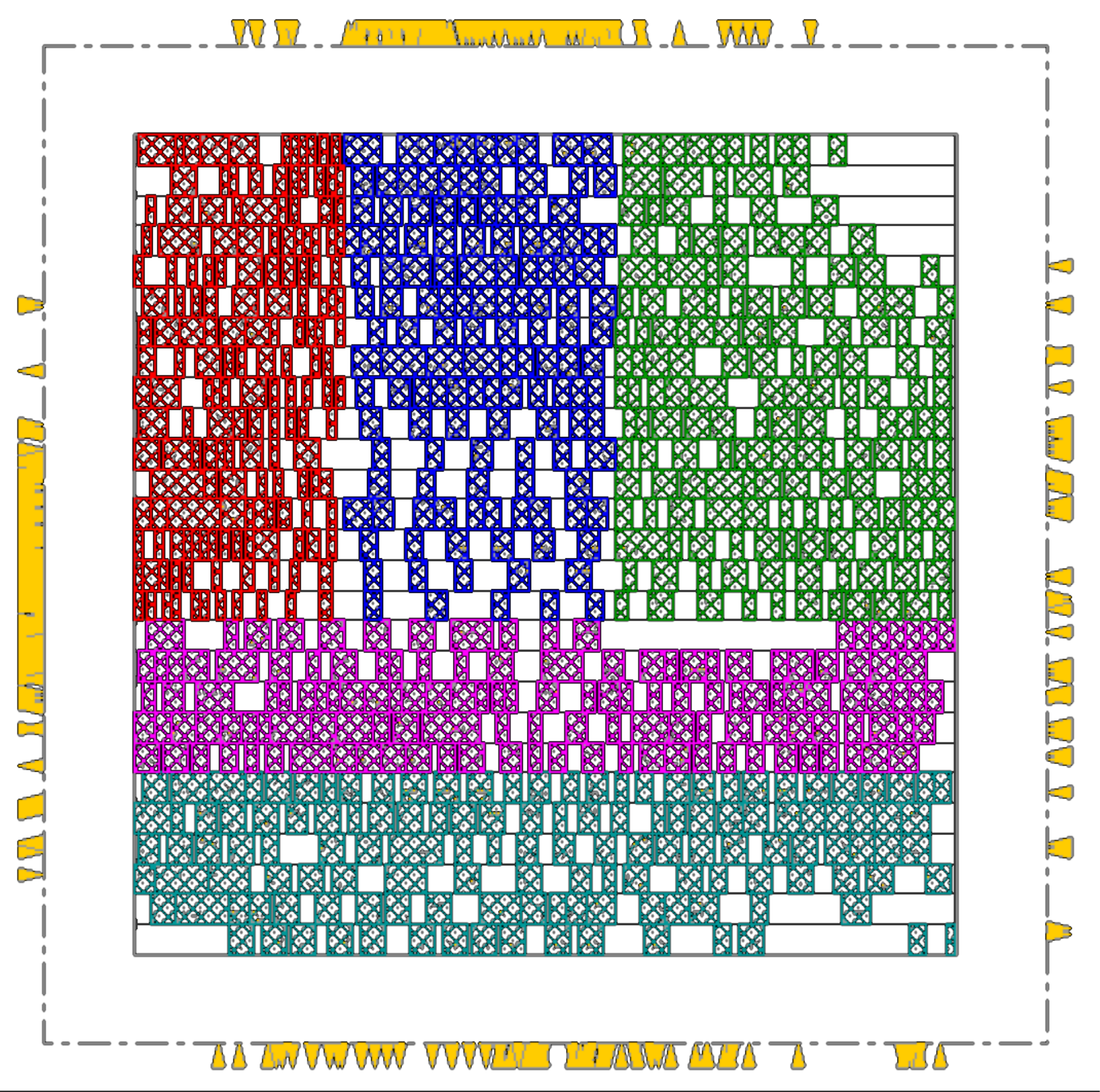}
 \label{fig:g-color_layout}}
 \subfloat[g-type2]{
\includegraphics[height=3.85cm]{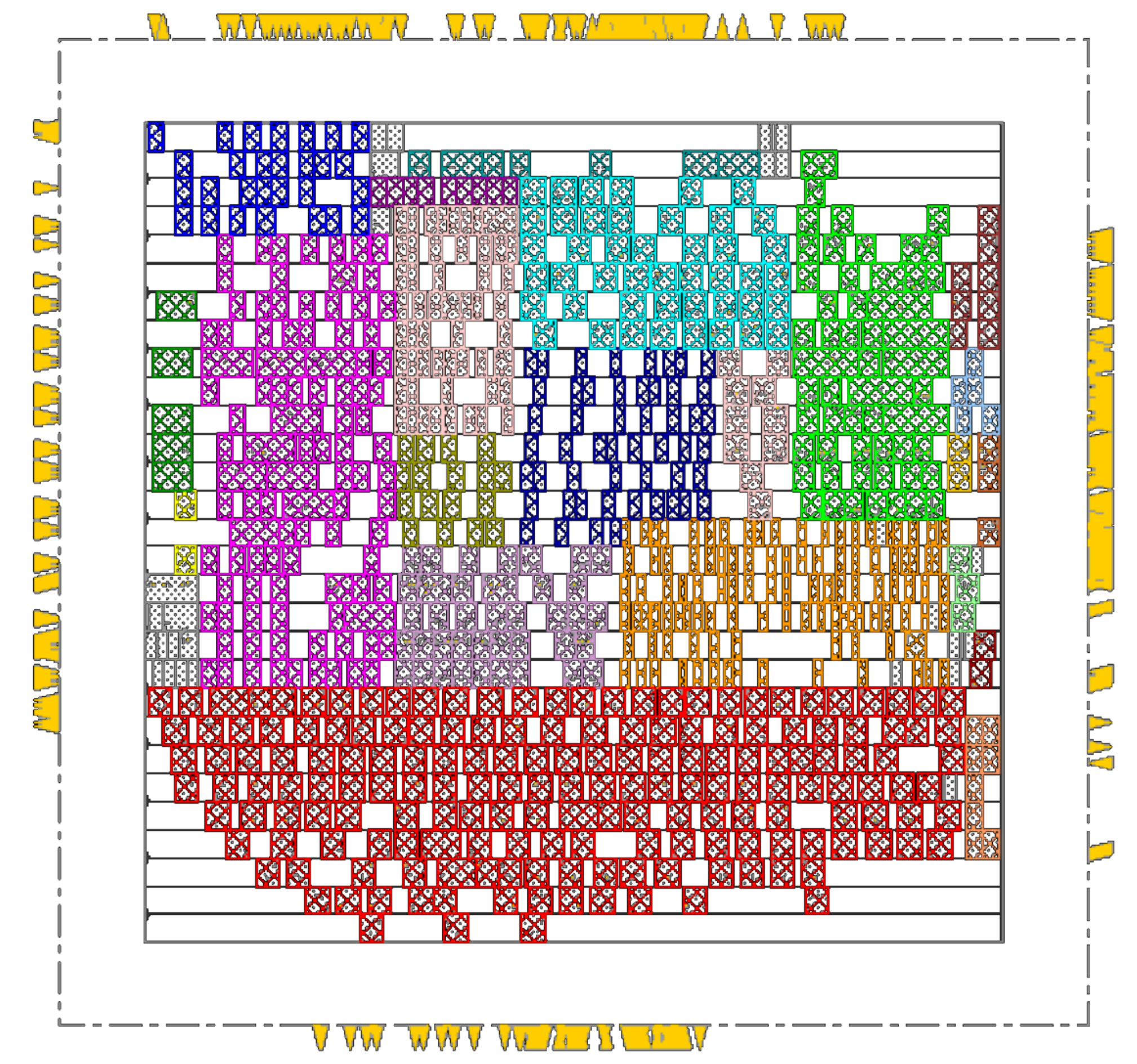}
 \label{fig:g-type2_layout}}
 \smallerspacecaption
\caption{Protected layouts for \emph{c7552}.
	Cells corresponding to particular clusters/partitions have identical colors.
}\label{fig:layout}
\end{figure}

Here
   we describe the
   steps for the layout generation utilizing our secure techniques.
   The steps
   are generic and can be easily embedded into any design flow.
As an example, two protected, cell-level layouts of \emph{c7552}
	are shown in Fig.~\ref{fig:layout}.

First,
we obtain the gate-level, technology-mapped netlist of the design to protect
(using the \emph{Cadence RTL compiler} along with the \emph{NanGate Open Cell Library}~\cite{nangate11}).
Next, we apply (one of) our proposed techniques
on that netlist
	to obtain the related design partitions.
Given these partitions, we generate a layout where all partitions are mapped to mutually exclusive layout regions called \emph{fences}.
A fence confines
the corresponding partition's cell placement within its boundaries.
This is an important
step
as it ensures that placement optimization cannot undermine the physical separation of cells dictated by partitions.
The actual system-level arrangement of all fences---which can be considered as floorplanning---is done automatically using \emph{Cadence Innovus}.
Finally, the layout is routed and finalized.
We like to emphasize the fact that we
target for a \emph{DRC-clean} layout by adapting the utilization target as needed.
We resolve any outstanding DRC issues, if any, and report the
power, performance, and area (PPA) numbers.

As the concept of split manufacturing hinges on two foundries,
   we split the DEF file into the FEOL and BEOL parts using a custom script.
   One aspect of our study is to investigate the cost-security implications at different split layers; we thus obtain multiple sets of FEOL and BEOL parts, for
   splitting from M1 up to M6.

\section{Experimental Results}\label{sec:results}

\begin{table}[b]
\centering
\tabsize
\caption{
	Benchmarks Selected 
from the ISCAS-85 and MCNC Suites, Along with Their Properties
}\label{tb:benchmarks}
\smallerspacecaption
\begin{tabular}{|c|c|c|c|}
\hline
 \textbf{Benchmark} & \textbf{Inputs} & \textbf{Outputs} & \textbf{Gate Count} \\ \hline \hline
apex2 & 39 & 3 & 610 \\ \hline
apex4 & 10 & 19 & 5,360 \\ \hline
c432 & 36 & 7 & 160 \\ \hline
c880 & 60 & 26 & 383 \\ \hline
c1908 & 33 & 25 & 880  \\ \hline
c2670 & 233 & 140 & 1,193 \\ \hline
c5315 & 178  & 123  & 2,307  \\ \hline
c7552 & 207 & 108 & 3,512  \\ \hline
des & 256 & 245 & 6,473  \\ \hline
ex1010 & 10 & 10 & 5,066  \\ \hline
\end{tabular}
\end{table}

\textbf{Setup:} Our secure layout techniques are implemented using \emph{Java OpenJDK 1.8.0\_121 64-bit Server VM}.
The layouts are generated using custom in-house scripts for \emph{Cadence Innovus 15.1} using the \emph{NanGate 45nm Open Cell Library}~\cite{nangate11} with ten metal layers.
Note that all metal layers are leveraged 
across all benchmarks for the sake of fair comparison.
The PPA analysis is
carried out at 0.95V for the slow process corner.
We evaluate the resilience of our protected layouts against the network-flow attack by Wang \emph{et al.}~\cite{jv-attack16}. For the latter, we run experiments on
an 8-core \emph{Intel Xeon i7-4790 CPU}, at 3.60GHz and with 16GB RAM. The operating system is \emph{Ubuntu 16.04.2 (xenial)}.
We conduct our experiments on the \emph{ISCAS'85} and \emph{MCNC} benchmark suites (Table~\ref{tb:benchmarks}).
While all those benchmarks are fully combinatorial, our techniques can be readily applied to sequential circuits as well.

\subsection{Security Analysis}

\begin{table}[tb]
\centering
\tabsize
\caption{Reduction in \emph{MI} (in \%) for the Proposed Techniques Compared to Original Layouts, When Split at M1}
\label{tb:mi}
\smallerspacecaption
\begin{tabular}{|c|c|c|c|c|}
\hline
 \textbf{Benchmark} & \textbf{Random} & \textbf{g-color} & \textbf{g-type1} & \textbf{g-type2} \\ \hline \hline
apex2 & 96.11 & 75.00 & 89.44 & 92.22 \\ \hline
apex4 & 96.67 & 90.00 & 96.67 & 93.33\\ \hline
c432 & 93.44 & 91.03 & 82.41 & 89.31\\ \hline
c880 & 96.84 & 88.42 & 86.84  & 89.47\\ \hline
c1908 & 95.00  & 79.29  & 85.71 & 85.71\\ \hline
c2670 & 97.22 & 85.56 & 89.44  & 94.44\\ \hline
c5315 & 98.00 & 92.00 & 94.00 & 92.00\\ \hline
c7552 & 98.89 & 91.11 & 90.00 & 88.89\\ \hline
des & 98.25 & 92.5 & 90.00  & 90.00\\ \hline
ex1010 & 96.67 & 93.33 & 93.33  & 93.33\\ \hline
\textbf{Avg.} & \textbf{96.61} & \textbf{87.43} & \textbf{87.33} & \textbf{89.56}\\ \hline
\end{tabular}
\end{table}

\textbf{Reduction in mutual information:}
The reduction in \emph{MI} for our different techniques, when compared to the original layouts, is presented in Table \ref{tb:mi}. 
As expected, random placement enables the largest reduction in \emph{MI} and, thus, presumably the best protection.
However, recall that this specific assessment is only applicable for splitting at M1. As for higher split layers,
one should rather consider the distance between remaining open pins/vias. These distances will be shorter on average, due to design tools (routers) seeking to shorten interconnects
wherever possible, thereby bringing the open pins/vias closer together (or even routing some of the nets already completely within the FEOL). Hence, the \emph{MI} as calculated for
splitting at M1 will become less expressive for higher layers.

\begin{figure*}[p]
 \centering
\subfloat[Split at M1]{
 \includegraphics[width=0.45\textwidth]{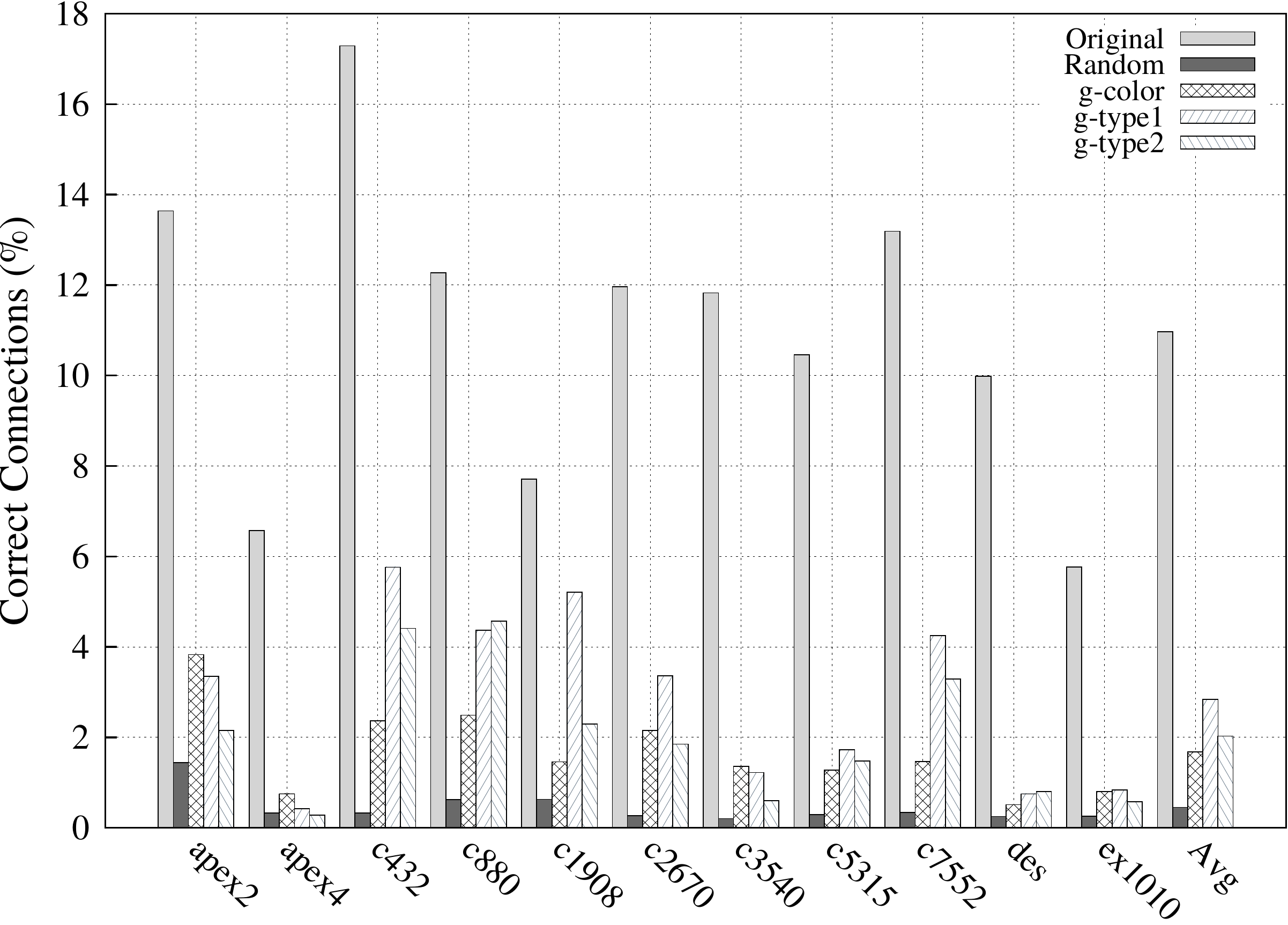}
 \label{fig:m1}}\quad
\subfloat[Split at M2]{
 \includegraphics[width=0.45\textwidth]{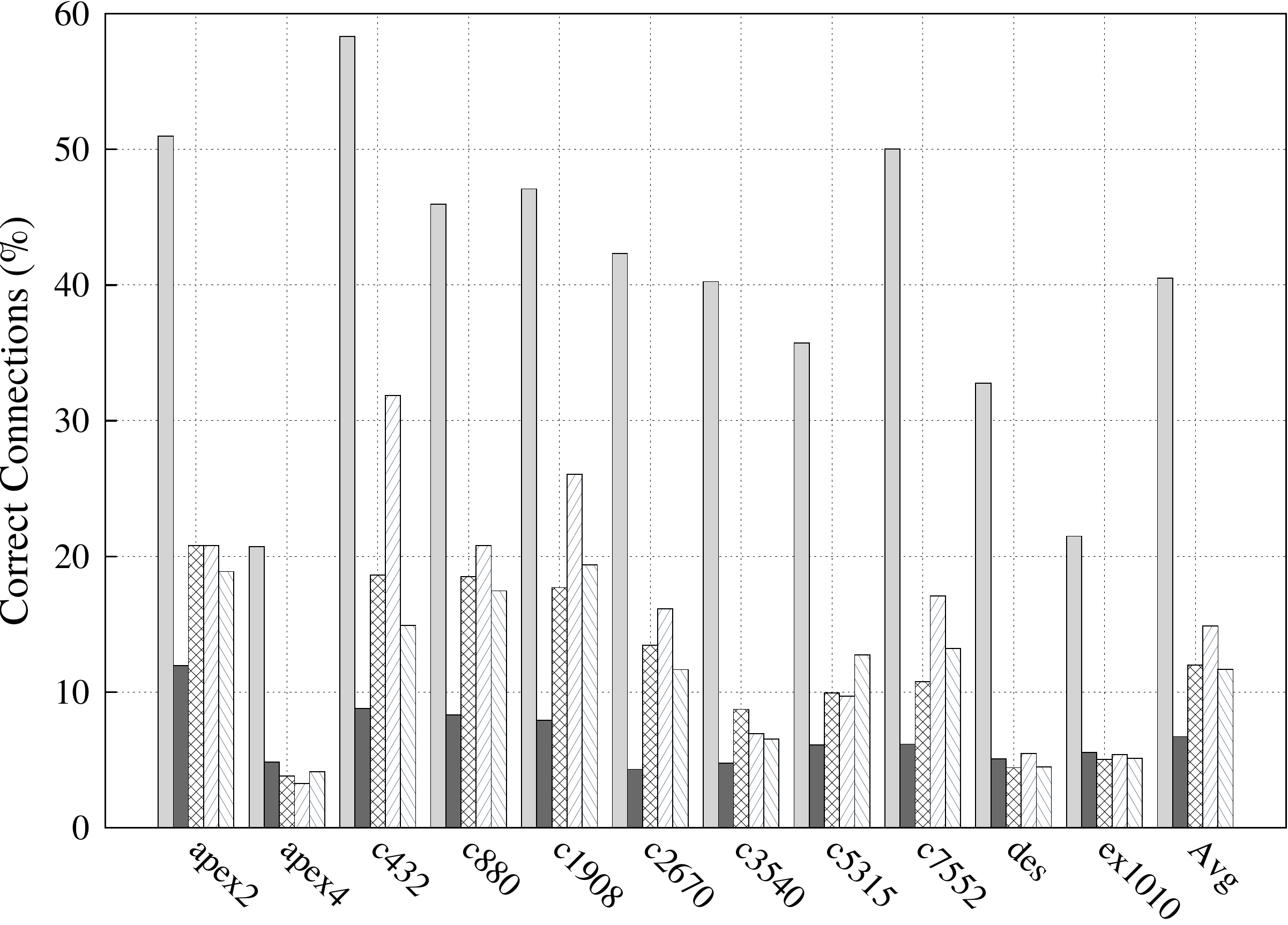}
 \label{fig:m2}}\\
\subfloat[Split at M3]{ 
 \includegraphics[width=0.45\textwidth]{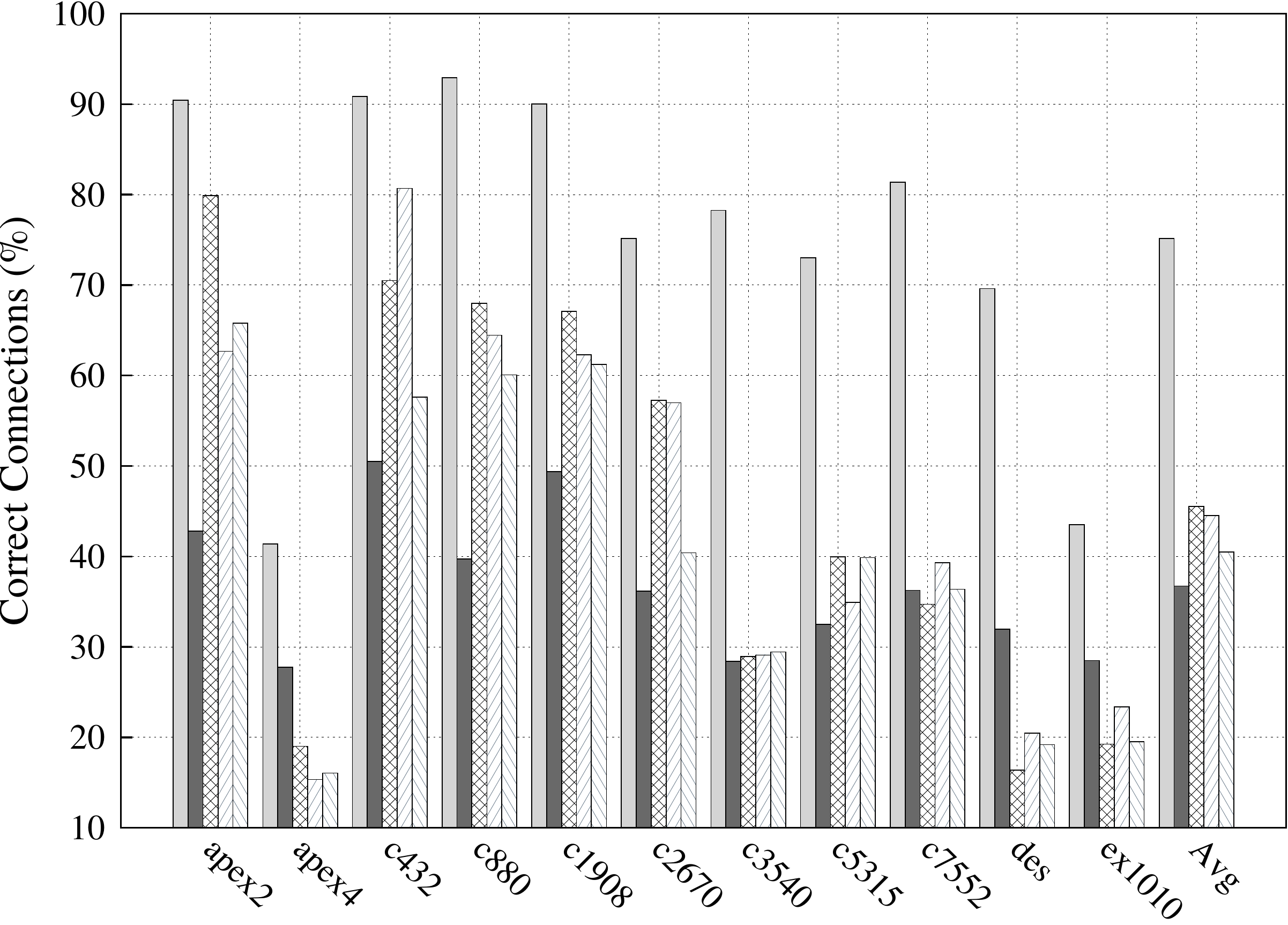}
 \label{fig:m3}}\quad
\subfloat[Split at M4]{
 \includegraphics[width=0.45\textwidth]{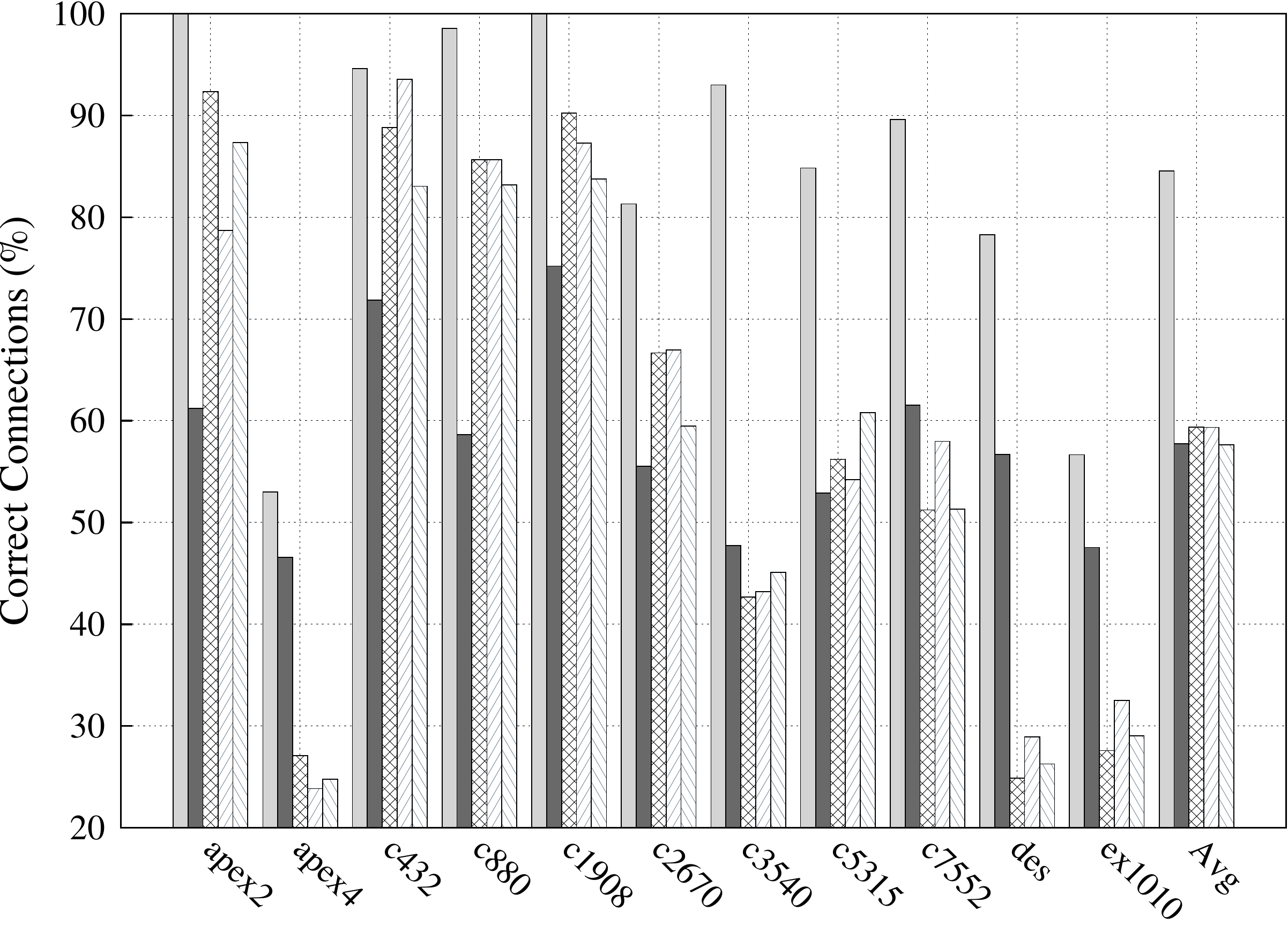}
 \label{fig:m4}}\\
\subfloat[Split at M5]{
 \includegraphics[width=0.45\textwidth]{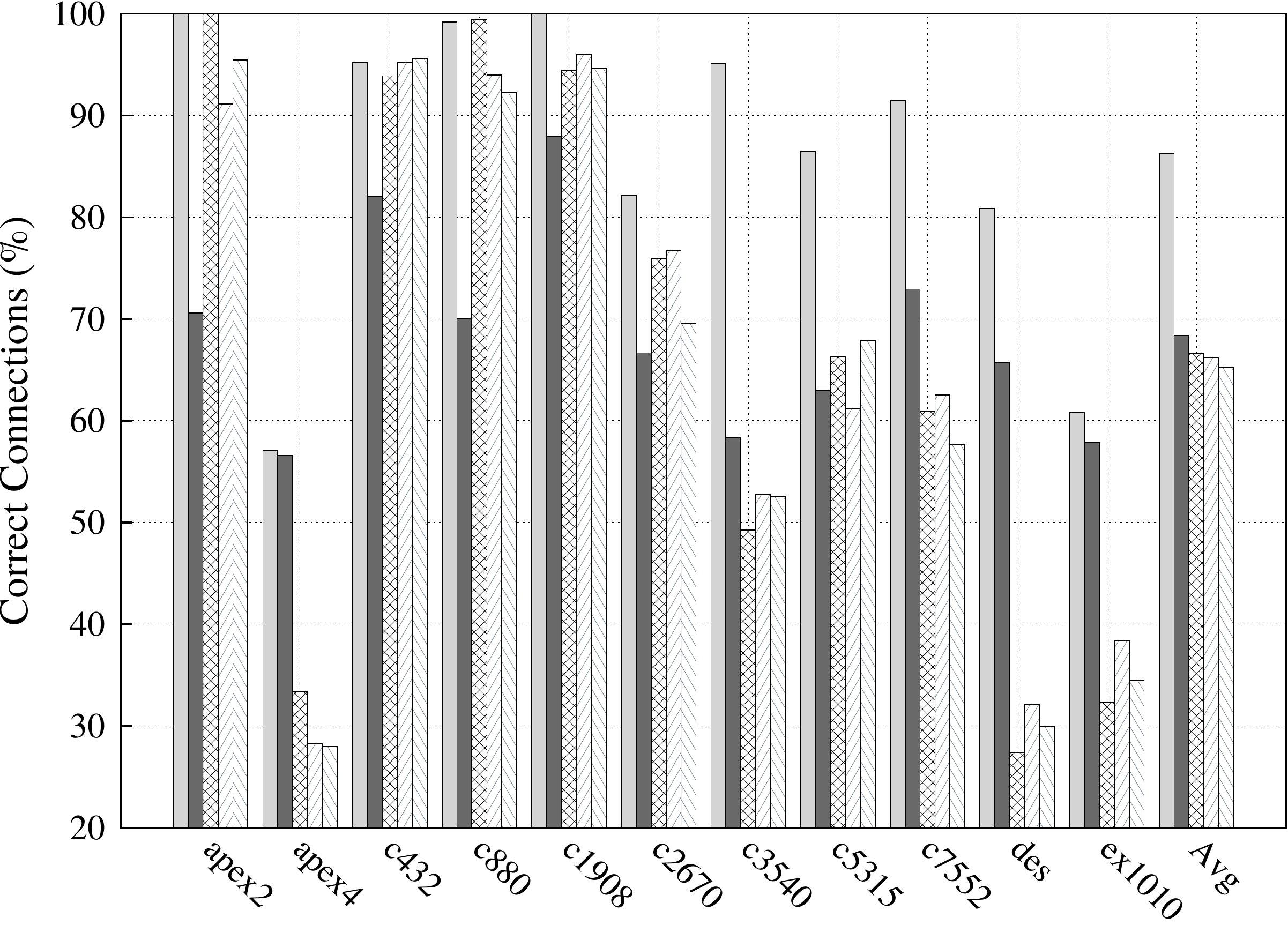}
 \label{fig:m5}}\quad
\subfloat[Split at M6]{ 
 \includegraphics[width=0.45\textwidth]{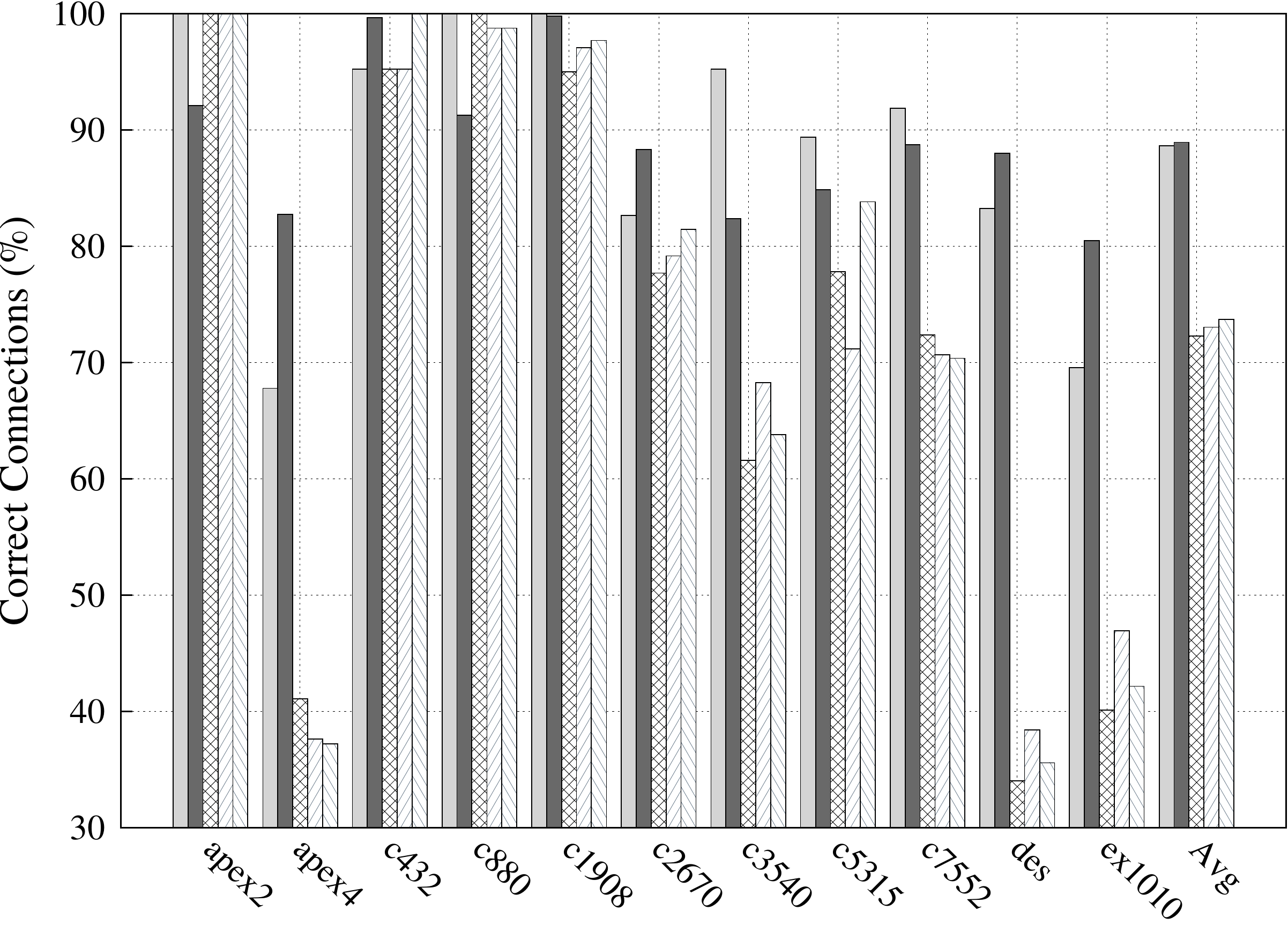}
 \label{fig:m6}}
 \smallerspacecaption
\caption{Correct connections (representing the attacker's success rates) for original and varyingly protected layouts, evaluated against the attack~\cite{jv-attack16} when split at different layers.}\label{fig:attack}
\end{figure*}

The above expectation---random placement is most secure, at least while splitting at lower layers---is corroborated while conducting the
attack~\cite{jv-attack16} across various split layers (Fig.~\ref{fig:attack}; see also below for further discussion).
While random placement is the most secure technique at lower split layers, it becomes less and less effective for higher layers, until the point (at M6) where even the original,
      unprotected layouts are more resilient.
In general, we observe the higher the split layer, the more
connections are correctly inferred and, thus, the
lower is the actual resilience.

\textbf{Resilience at M1:} It is evident from Fig.~\ref{fig:m1} that
only a few connections are recovered correctly across all benchmarks. Thus, the resilience when splitting at M1 is generally high.
Still, as expected, we find that the original layouts are easiest to attack.
This reiterates the fact that 
design tools shall be
reinforced with the help of security metrics (such as \emph{MI}) once split manufacturing is considered.

We observe that
	randomization enables the highest resilience.
This is in agreement with our findings above, i.e.,
randomization achieves the largest reduction in \emph{MI}.
Again, this is expected as we ``dissolve'' the hints of connectivity by randomly perturbing the placement of all gates.
Unfortunately, randomization comes at a hefty cost for PPA (up to 600\%); see also Section~\ref{sec:ppa}.

Our proposed partitioning techniques perform quite well for security; we are able to significantly reduce the percentage of correct connections when compared to original layouts.
In fact, we observe on average reductions of
    6.54$\times$, 3.86$\times$, and 5.41$\times$ for g-color, g-type1, and g-type2, respectively.
    As illustrated in Fig.~\ref{fig:m1} (and Fig.~\ref{fig:cc_mi_wl}), we can achieve similar resilience
    when compared to randomization (with lower wirelength overheads at the same time).

\textbf{Resilience at M2 and M3:}
As expected, the resilience generally decreases across all techniques and benchmarks when compared to M1
(Figs.~\ref{fig:m2} and~\ref{fig:m3}).
Interestingly enough,
the advances of our techniques
still carry over to a great extent.
On average, we reduce the correct connections by 2.73--3.47$\times$ and 1.64--1.85$\times$ while
splitting at M2 and M3, respectively, as opposed to original layouts. 
Moreover, for relatively large benchmarks under consideration (i.e., \emph{apex4}, \emph{des}, and \emph{ex1010}), our techniques are even on a par with randomization.

\textbf{Resilience at M4 and above:} Once we split at layer M4 or above, we still achieve average reductions by 1.4$\times$, 1.3$\times$, and 1.2$\times$
over
original layouts
(Figs.~\ref{fig:m4}, \ref{fig:m5}, and~\ref{fig:m6}).
For relatively large benchmarks (\emph{apex4}, \emph{des}, and \emph{ex1010}), we even achieve reductions by 1.52--1.75$\times$.
Also note that our
techniques are on average on a par with layout randomization for M4 and M5, and notably excel it for M6, by 1.2$\times$.
This clearly indicates that only thoughtful placement-centric protection schemes can imply some resilience
also for higher split layers and relatively large benchmarks.

\begin{figure}[tb]
\centering
\includegraphics[width=0.85\textwidth]{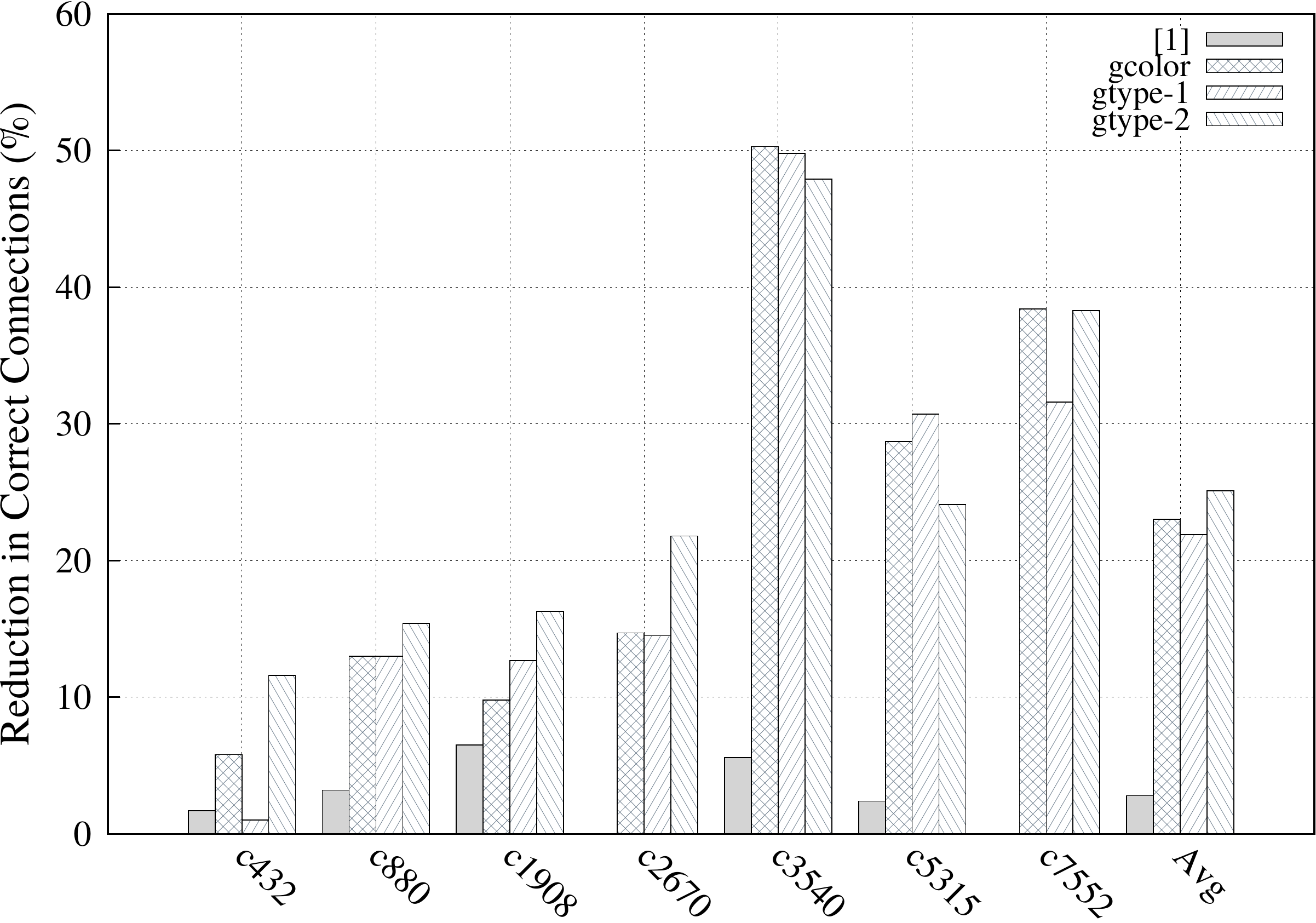} 
\caption{Reduction in correct connections achieved for the protection scheme in~\cite{jv-attack16} and our techniques, for splitting at M4.
	The data for~\cite{jv-attack16} is quoted from their recent publication~\cite{wang17} which is also based on the same attack as in~\cite{jv-attack16}.
}
\smallerspacecaption
\label{fig:comparison}
\end{figure}

\textbf{Comparison with Wang \emph{et al.}~\cite{jv-attack16}:} We compare our work to the most recent in \emph{placement-centric protection} by Wang~\emph{et
	al.}~\cite{jv-attack16} in Fig.~\ref{fig:comparison}.
Note that we compare for splitting at M4 since the layouts provided to us indicate this split layer.
We lower the number of correct connections on average by 21.9--25.1\% when compared to the original layouts---this is an
improvement of $\approx$8$\times$ over~\cite{jv-attack16}.

Besides that, we cannot directly compare with other studies such as~\cite{magana16,wang17}; these are 
\emph{routing-centric protection} techniques and we are also not made aware of all essential details of their protected layouts (such as the technology files).
However, the data presented in~\cite{wang17} indicates that our techniques are still competitive.

\subsection{Layout-Level Cost Analysis}\label{sec:ppa}
\begin{figure}[!tb]
 \centering
 \subfloat{
 \includegraphics[width=0.84\textwidth]{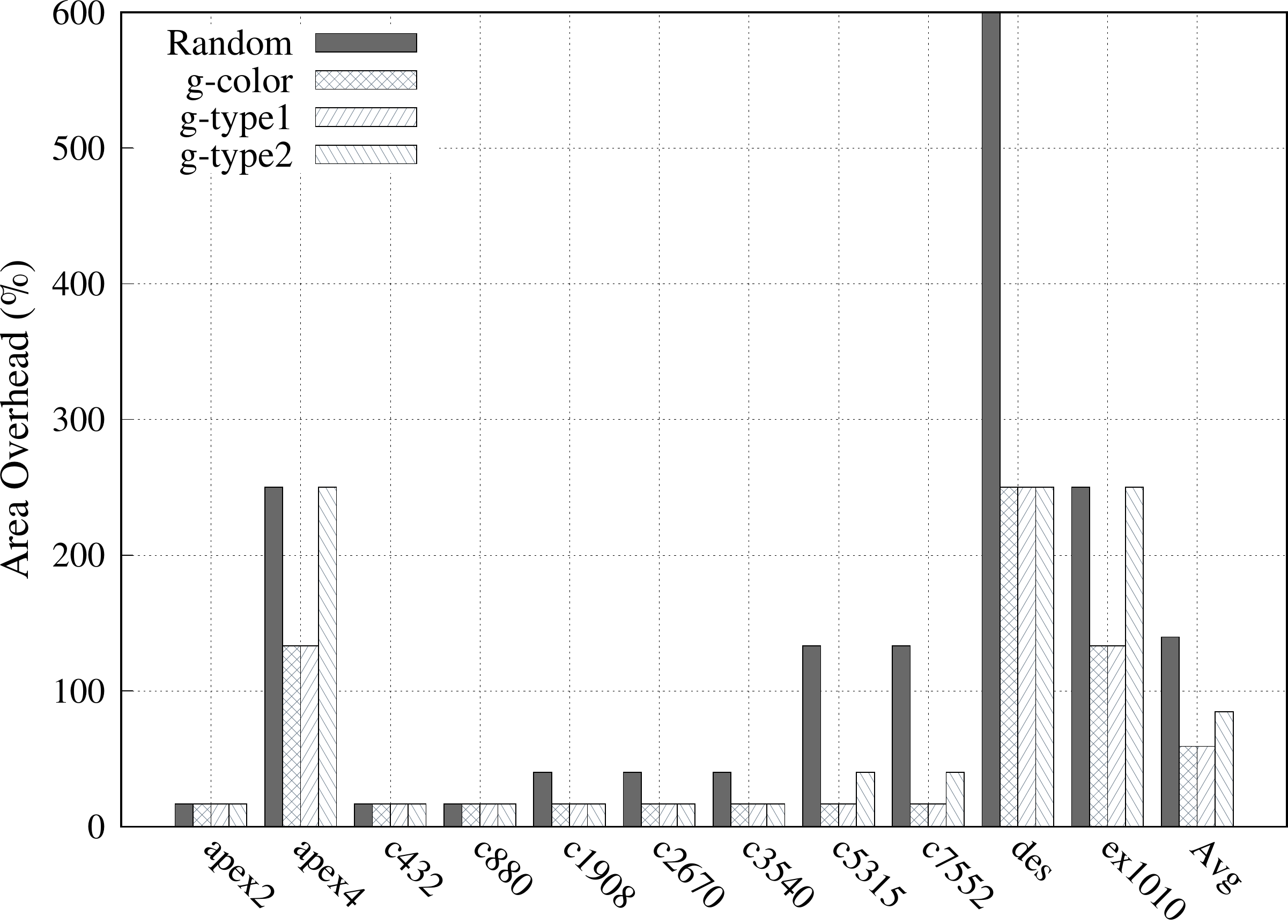}
 \label{fig:area}}\\
	 \vspace{-2.5mm}
 \subfloat{
 \includegraphics[width=0.84\textwidth]{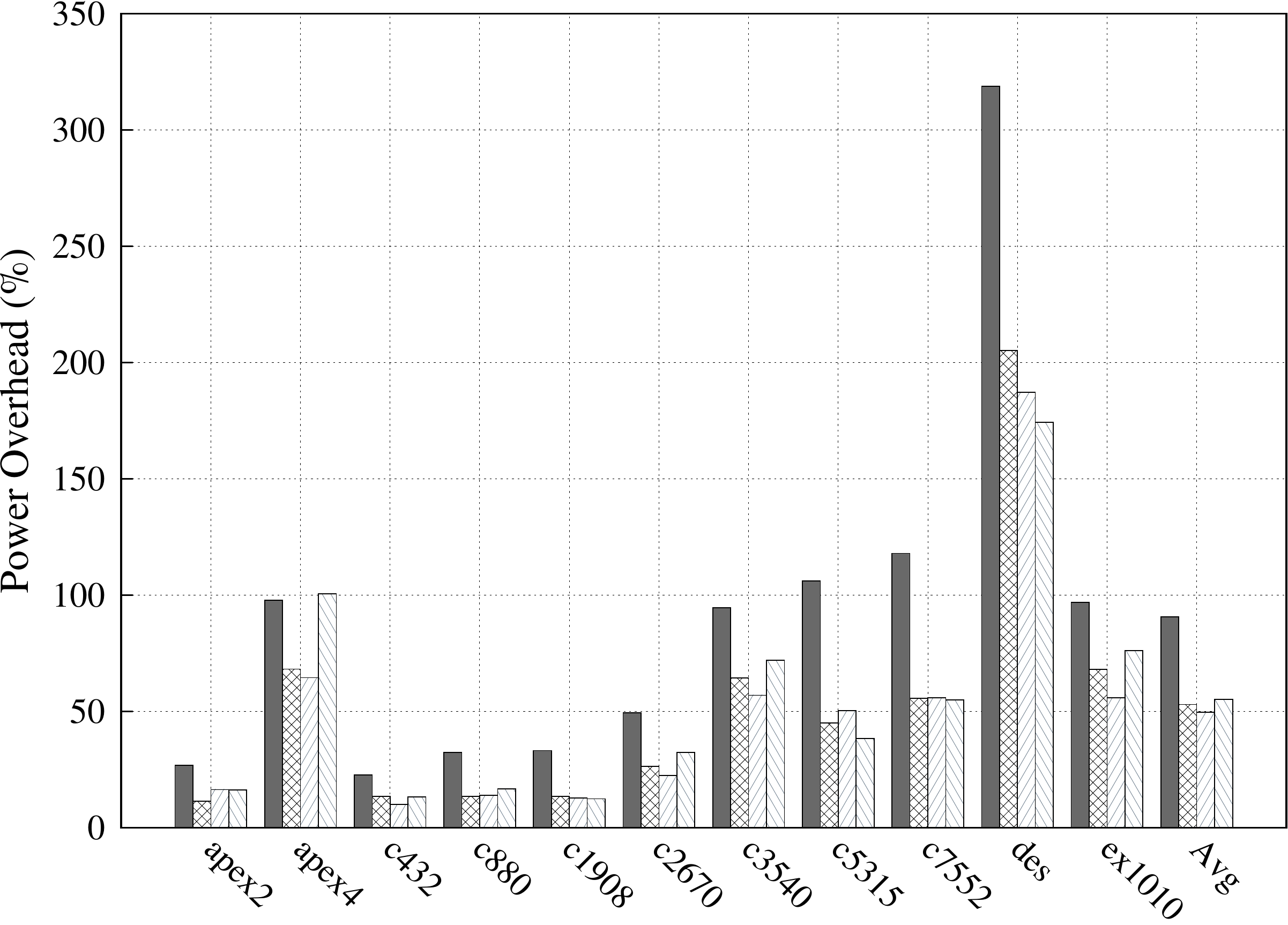}
 \label{fig:power}}\\
	 \vspace{-2.5mm}
 \subfloat{
\includegraphics[width=0.84\textwidth]{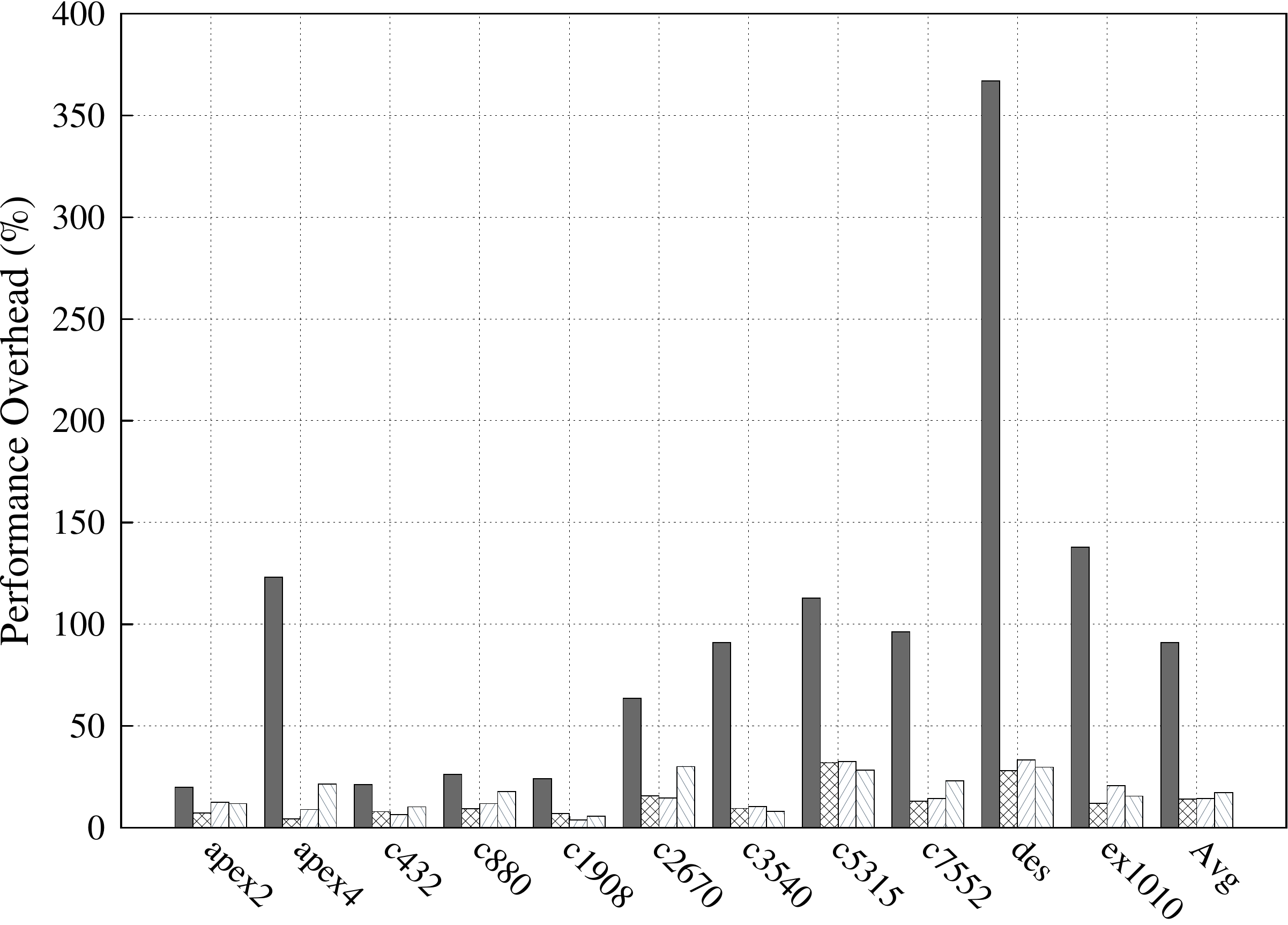}
 \label{fig:delay}}
 \smallerspacecaption
\caption{Layout overheads for various techniques in contrast to the original, unprotected layouts.}\label{fig:overhead}
\end{figure}

\textbf{Area
		overheads:}
Recall that we adapt the utilization rates as needed to enable DRC-clean layouts; the reported area cost accordingly captures the effect of upscaling die outlines.
While layout randomization enables the most resilient layouts on average, it incurs prohibitive overheads (Fig.~\ref{fig:overhead}).
We note that area cost scales up
significantly for relatively large benchmarks under consideration (i.e., \emph{apex4}, \emph{des}, and \emph{ex1010}); however, these benchmarks are still decent in size when compared to
state-of-the-art industrial designs. Hence, randomizing layouts is not scalable.
In contrast, we observe
that our techniques g-color and g-type1 induce on average 60\% area cost.
Applying g-type2, however, results in larger overhead, sometimes comparable to randomization.
Since g-type2 induces on average more partitions, the system-level routing for those partitions becomes more challenging and congested, which can only be managed by larger die
outlines.
It is easy to see that routability poses a major challenge for any protection scheme ``dissolving'' the connectivity of gates and their placement.
Naturally, a larger die outline also lengthens wires to some degree, which \emph{may} also impact power and performance.

\textbf{Power and performance overheads:}
As for layout randomization, both the power
and delay overhead are prohibitive.
Again, recall that randomization deliberately and uncontrollably ``rips apart'' connected gates. Even sophisticated design optimization in later stages (timing-driven routing, 
		clock gating, etc., see also~\cite{KLMH11}) may handle the related overhead only to some degree.

As for our novel layout techniques, we obtain significantly lower overheads.
We observe average power overheads of 50\% across all benchmarks, which is an improvement of 1.6$\times$ over layout randomization.
Further, we observe average delay overheads less than 18\% in all the benchmarks under consideration; this translates to an improvement of 5$\times$ over randomization.

\textbf{Comparison with Wang \emph{et al.}~\cite{jv-attack16}:} While the respective layouts are available to us, we have not been made aware of the technology files and the
specifics of the physical-design
setup.
Hence, we cannot reasonably contrast the PPA cost at the layout level.
Further, PPA cost is also omitted in~\cite{jv-attack16} itself.
While the wirelength numbers reported in~\cite{jv-attack16} may be lower,
it is well known that wirelength is only one aspect among many others to impact PPA cost~\cite{shelar13}.

\subsection{Discussion}
\label{sec:discussion}
\textbf{Impact of technology libraries:}
We also investigate the impact of different technology libraries on both security and cost when using our placement-centric techniques.
For these experiments, we additionally synthesized the \emph{c7552} circuit using a library constrained to the three essential gates: NAND, NOR, and inverters.

First, note that
g-color is agnostic with respect to the library; the partitioning is solely dictated by connectivity.
For g-type, however, there is an interdependency between the library and the final layout since partitions are based on gates of the same type.
We observe a
direct relation between protection and the number of partitions (i.e., gates in the library): the fewer partitions, the lower the resilience (in terms of more correctly recovered
		connections, see Table~\ref{tb:condense}).
Besides the adverse impact on security, it should be noted that such a significantly constrained library is not practical
   as it offers very little room for design optimization.
In short, an enriched library not only offers significantly more room for design optimization, but it also enables higher resilience while using g-type.\footnote{Depending on both the
	design to protect and the library, there may be cases where only a few gates
		remain within a partition. This might undermine the resilience of g-type to some
		degree, as the arrangement of very small partitions may leak their underlying connectivity.
As a countermeasure, these partitions could be balanced by adding dummy gates as needed. Note that this would also 
prevent leaking the functional composition of the design~\cite{otero15,jaggu14}.}

\begin{table}[tb]
\centering
\tabsize
\caption{Correct Connections (in \%) for \emph{c7552}, Protected using g-type, and
for Different Libraries and Split Layers}
\label{tb:condense}
\smallerspacecaption
\begin{tabular}{|c|c|c|c|c|}
\hline
\multirow{2}{*}{\textbf{Split layer}}
&\multicolumn{2}{c|}{\textbf{Full library}}
&\multicolumn{2}{c|}{\textbf{Constrained library}} \\
\cline{2-5}
& \textbf{g-type1} & \textbf{g-type2}
& \textbf{g-type1} & \textbf{g-type2} \\
	\hline \hline
M1 & 4.3 & 3.3 & 16.1 & 13.9 \\ \hline
M2 & 17.1 & 13.2 & 45.5 & 37.8 \\ \hline
M3 & 39.3 & 36.4 & 59.5 & 53.7 \\ \hline
M4 & 57.9 & 51.3 & 75.5 & 69.3 \\ \hline
M5 & 62.5 & 57.7 & 76.9 & 71.4 \\ \hline
M6 & 70.7 & 70.4 & 89.6 & 85.6 \\ \hline
\end{tabular}
\end{table}

\textbf{Trade-off for cost and security:}
We note that the layout resilience varies greatly across the different split layers and different benchmarks.
While splitting at M1, M2, or even M3 still offers a reasonable protection also for relatively small benchmarks (which are easier to attack in general), splitting at M4 or above can
only protect relatively large benchmarks in comparison.
In general, splitting at higher layers implies lower commercial cost, since the trusted BEOL fab is then only required to handle few metal layers having relatively large
pitches~\cite{jv-attack13,iarpa}.

Determining the design-specific ``sweet spot'' for cost and security is thus an essential challenge for split manufacturing.
Towards this end,
we advocate our metric (mutual information) as another design criteria for future, security-aware tools.
Moreover, we like to emphasize the fact that only thoughtful placement-centric schemes like ours (and unlike layout randomization) can provide some degree of protection at higher
split layers as well.

\textbf{Towards better protection at higher split layers:}
   While our techniques already provide comparable protection to randomization at lower layers and even translate to better protection at higher layers, we still observe the
   general trend of increasingly successful recovery once the attack targets at higher layers.
   Thus, an interesting question arising is whether one can further strengthen our placement-centric schemes also for higher layers.
   We believe that this requires applying both placement- and routing-centric techniques in conjunction; this will be the scope of future work.

\section{Conclusion}\label{sec:concl}
In this work, we first formulate an information-theoretic metric---the mutual information between the connectivity and distances of gates---which helps to analyze the protection of
physical layouts against proximity attacks.
Our metric
can measure the security in an objective and efficient way, as compared to empirical and attack-based evaluation schemes.  We show further that randomizing the layout/placement can
reduce the mutual information, but only at an excessive overhead. Thus, we also present two effective, placement-centric techniques (namely, g-color and g-type) which enable
competitive (sometimes even superior) protection along with an acceptable layout cost.
For future work, we plan to extend our approach towards protection at both the FEOL \emph{and} BEOL end.

\section*{Acknowledgments}
The authors are grateful to Tri Cao and Jeyavijayan (JV) Rajendran (University of Texas at Dallas)
for providing their network-flow attack and their protected layouts of~\cite{jv-attack16}.

This work was supported in part by the National Science Foundation (NSF) under Grant 1553419,
     the Computing and Communication Foundations (NSF/CCF) under Grant 1319841,
     a grant from the Semiconductor Research Corporation (SRC),
     and the New York University/New York University Abu Dhabi (NYU/NYUAD) Center for Cyber Security (CCS).
Any views expressed are the authors' own and do not necessarily reflect the views of the NSF or SRC.

\newcommand{\BIBdecl}{\setlength{\itemsep}{0.03em}}
\bibliographystyle{IEEEtran}
\bibliography{main}

\end{document}